%% file: IFIP.tex
\documentclass[sigconf]{acmart}

\usepackage{booktabs} 
\usepackage{enumitem}
\usepackage{amsmath}

\usepackage{amssymb}

\usepackage{fixmath}

\usepackage{algorithm}
\usepackage{algorithmic}

\setcopyright{rightsretained}

\acmDOI{10.475/123_4}

\acmISBN{123-4567-24-567/08/06}

\acmConference[]{ACM IFIP WG 7.3 Performance}{2017}{New York}
\acmYear{2017}
\copyrightyear{2017}

\acmPrice{15.00}

\begin{document}
\title{GB-PANDAS: Throughput and heavy-traffic optimality analysis for affinity scheduling}

\author{Ali Yekkehkhany}
\affiliation{
\institution{University of Illinois at Urbana-Champaign}
}
\author{Avesta Hojjati}
\affiliation{
\institution{University of Illinois at Urbana-Champaign}
}
\author{Mohammad H Hajiesmaili}
\affiliation{
\institution{Johns Hopkins University}
}

\begin{abstract}
Dynamic affinity scheduling has been an open problem for nearly three decades. The problem is to dynamically schedule multi-type tasks to multi-skilled servers such that the resulting queueing system is both stable in the capacity region (throughput optimality) and the mean delay of tasks is minimized at high loads near the boundary of the capacity region (heavy-traffic optimality).
As for applications, data-intensive analytics like MapReduce, Hadoop, and Dryad fit into this setting, where the set of servers is heterogeneous for different task types, so the pair of task type and server determines the processing rate of the task. The load balancing algorithm used in such frameworks is an example of affinity scheduling which is desired to be both robust and delay optimal at high loads when hot-spots occur.
Fluid model planning, the MaxWeight algorithm, and the generalized $c\mu$-rule are among the first algorithms proposed for affinity scheduling that have theoretical guarantees on being optimal in different senses, which will be discussed in the related work section. All these algorithms are not practical for use in data center applications because of their non-realistic assumptions. The join-the-shortest-queue-MaxWeight (JSQ-MaxWeight), JSQ-Priority, and weighted-workload algorithms are examples of load balancing policies for systems with two and three levels of data locality with a rack structure. In this work, we propose the Generalized-Balanced-Pandas algorithm (GB-PANDAS) for a system with multiple levels of data locality and prove its throughput optimality. We prove this result under an arbitrary distribution for service times, whereas most previous theoretical work assumes geometric distribution for service times. The extensive simulation results show that the GB-PANDAS algorithm alleviates the mean delay and has a better performance than the JSQ-MaxWeight algorithm by up to twofold at high loads.
We believe that the GB-PANDAS algorithm is heavy-traffic optimal in a larger region than JSQ-MaxWeight, which is an interesting problem for future work.
\end{abstract}

%
%



\keywords{Affinity scheduling, MapReduce, near-data scheduling, data locality, data center, big data.}

\maketitle

\input{samplebody-conf}

\end{document}

%% file: samplebody-conf.tex
\section{Introduction}

Affinity scheduling refers to the allocation of computing tasks on computing nodes in an efficient way to minimize a cost function, for example the mean task completion time \cite{padua2011encyclopedia}. The challenge of how to load balance the tasks between the computing nodes made this problem open for nearly three decades. More specifically, a computing node can have different speeds for different task types, which is referred to as different levels of data locality in the context of data center load balancing. As a result, a dilemma between throughput and performance emerges in the affinity scheduling problem.

The works by Harrison and Lopez \cite{harrison1998heavy, harrison1999heavy}, and Bell and Williams \cite{bell2001dynamic, bell2005dynamic} on affinity scheduling require the arrival rates of all task types in addition to existence of one queue per task type, which are not realistic assumptions for applications like load balancing for data centers. On the other hand, even though the generalized c$\mu$-rule algorithm by Stolyar and Mandelbaum \cite{stolyar2004maxweight, mandelbaum2004scheduling} does not use the arrival rates of task types, it still requires one queue per task type, which makes the system structure complicated. Moreover, it does not minimize the mean task completion time. Even though the recent works by Wang et al. on JSQ-MaxWeight \cite{wang2016maptask} and Xie et al. on the JSQ-Priority and Weighted-Workload algorithms \cite{xie2015priority, xie2016scheduling, xie2016schedulingPhD} resolve the above issues; however, they focus on a special case of affinity scheduling for data centers with two or three levels of data locality, where the service time of a computing node has geometric distribution.

\begin{figure*}[t]
\centering
\includegraphics[scale=0.44]{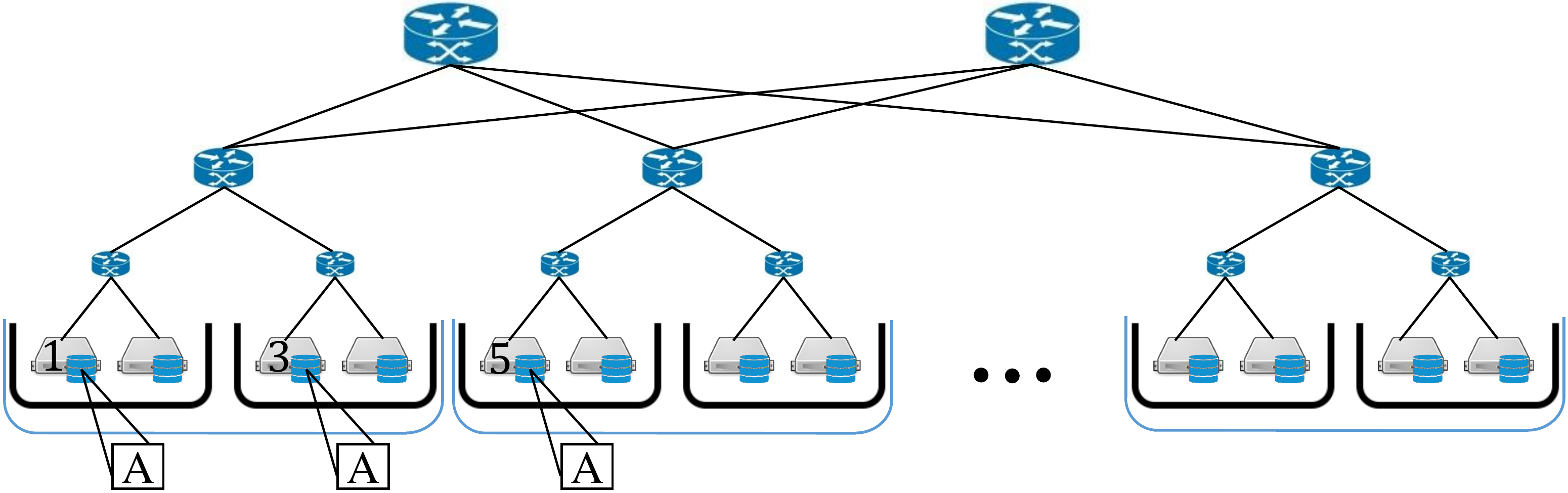}
\caption{A typical data center architecture with four levels of data locality.}
\label{SystemModel}
\end{figure*}

In this work, we propose the Generalized-Balanced-Priority-Algorithm-for-Near-Data-Scheduling (Generalized-Balanced-Pandas or GB-PANDAS) with a new queueing structure for the affinity scheduling problem. The GB-PANDAS algorithm does not require the arrival rates of task types and is for the general case with multiple levels of data locality. We establish the capacity region of the system for affinity scheduling problem and prove the throughput optimality of our proposed algorithm. The service times are assumed to be non-preemptive and they can have an arbitrary distribution, not necessarily geometric distribution which is the main assumption in \cite{xie2015priority, xie2016scheduling}, so we have to use a different Lyapunov function than the ordinary sum of cubic of the queue lengths to prove the throughput optimality of the GB-PANDAS algorithm. We take the map task scheduling problem, which is described in the rest of this section, as a platform to test the performance of our algorithm versus the state-of-the-art algorithms that are either widely used in the industry or have theoretical guarantees for optimality in some senses. The extensive simulation results show that our proposed algorithm performs better than other algorithms at heavy-traffic loads. We think that the GB-PANDAS algorithm is heavy-traffic optimal in a larger region than the JSQ-MaxWeight algorithm proposed by Wang et al. \cite{wang2016maptask}. For future work, one can start with the simpler problem of proving that GB-PANDAS is heavy-traffic optimal in the same region where JSQ-MaxWeight is heavy-traffic optimal. In the following, a short but inclusive review on the applications of the affinity scheduling problem in data centers and different venues of research in this field is presented.

\subsection{Applications of Affinity Scheduling in MapReduce Framework}

In large scale data-intensive applications like the health-care industry, ad placement, online social networks, large-scale data mining, machine learning, search engines, and web indexing, the de facto standard is the MapReduce framework. MapReduce framework is implemented on tens of thousands of machines (servers) in systems like Google's MapReduce \cite{dean2008mapreduce}, Hadoop \cite{apache}, and Dryad \cite{isard2007dryad} as well as grid-computing environments \cite{isard2009quincy}. Such vast investments do require improvements in the performance of MapReduce, which gives them new opportunities to optimize and develop their products faster \cite{ananthanarayanan2011scarlett}.
In MapReduce framework, a large data-set is split into small data chunks (typically 64 or 128 MB) and each one is saved on a number of machines (three machines by default) which are chosen uniformly at random. A request for processing the large data-set, called a job, consists mainly of two phases, map and reduce. The map tasks read their corresponding data chunks which are distributed across machines and output intermediary key-value results. The reduce tasks aggregate the intermediary results produced by map tasks to generate the job's final result.

In MapReduce framework, a master node (centralized scheduler) assigns map and reduce tasks to slaves (servers) in response to heartbeats received from slaves. Since jobs are either map-intensive or only require map tasks \cite{kavulya2010analysis, chen2012energy}, and since map tasks read a large volume of data, we only focus on map task scheduling as an immediate application of our load balancing algorithm. Local servers of a map task refer to those servers having the data associated with the map task. Local servers process map tasks faster, so the map tasks are preferred to be co-located with their data chunks or at least be assigned to machines that are close to map tasks' data, which is commonly referred to as near-data scheduling or scheduling with data locality.

In contrast to the improvements in the speed of data center networks, there is still a huge difference between accessing data locally and fetching it from another server \cite{ananthanarayanan2012pacman, xie2016pandas}. Hence, improving data locality increases system throughput, alleviates network congestion due to less data transmission, and enhances users' satisfaction due to less delay in receiving their job's response. There are mainly two approaches to increase data locality: 1) Employing data replication algorithms to determine the number of data chunk replicas and where to place them (instead of choosing a fixed number of machines uniformly at random, which is done in Google File System \cite{sanjay2003google} and Hadoop Distributed File System \cite{apache}). For more details see the algorithms Scarlett \cite{ananthanarayanan2011scarlett} and Dare \cite{abad2011dare}. 2) Scheduling map tasks on or close to local servers in a way to keep balance between data-locality and load-balancing (assigning all tasks to their local machines can lead to hotspots on servers with popular data). These two methods are complementary and orthogonal to each other. The focus of this paper is on the second method.

In addition to data-locality, fairness is another concern in task scheduling which actually conflicts with data-locality. Hence, in this work we only focus on data-locality and do not elaborate on fair scheduling. Note that our proposed scheduling algorithm can cooperate with job-level fair scheduling strategies to partly achieve fairness, which will be illustrated further in Section \ref{SystemModelSection} (as mentioned, both complete fairness and data-locality cannot be obtained at the same time). For more details on fair scheduling see the algorithms Delay Scheduling \cite{zaharia2010delay} and Quibcy \cite{isard2009quincy}.


The rest of the paper is outlined as follows. Section \ref{SystemModelSection} describes the system model and quantifies an outer bound of the capacity region (this outer bound is shown in Section \ref{ThroughputOptimalitySection} to be the capacity region of the system). Section \ref{BP} presents the GB-PANDAS algorithm and the queue dynamics under this algorithm. Section \ref{ThroughputOptimalitySection} analyzes the throughput optimality of our algorithm and Section \ref{SimulationResultsSection} evaluates the performance of the algorithm. Section \ref{RelatedWorkSection} discusses related work and Section \ref{ConclutionSection} concludes the paper with a discussion of opportunities for future work.

\section{System Model}
\label{SystemModelSection}
A discrete time model for the system is studied, where time is indexed by $t \in \mathbb{N}$.
The system consists of $M$ servers indexed by $1, 2, \cdots, M$. Let $\mathcal{M} = \{1, 2, \cdots, M\}$ be the set of servers. In today's typical data center architecture, these servers are connected to each other through different levels of switches or routers. A typical data center architecture is shown in Figure \ref{SystemModel}, which consists of servers, racks, super racks, top of the rack switches, top of the super rack switches, and core switches. \\
Remark. Note that our theoretical analysis does not care about the rack structure in data centers, so the result of throughput optimality of the GB-PANDAS algorithm is proved for an arbitrary system with $N$ levels of data locality (as an example, recall the affinity scheduling problem). The rack structure is only proposed as an incentive for this theoretical work, but the result is more general.


Considering the MapReduce framework for processing large data-sets, the data-set is split into small data chunks (typically of size 128 MB), and the data chunks are replicated on $d$ servers where the default for Hadoop is $d = 3$ servers. The bottleneck in MapReduce is the Map tasks, not the Reduce task, so we only consider Map tasks in this paper. \\
\textbf{Task Type:} In the Map stage, each task is associated to the processing of a data chunk, and by convention we denote the type of the task by the label of the three servers where the data chunk is stored \cite{hadoop, xie2016scheduling}. As an example, the task associated to processing data chunk $A$ shown in Figure \ref{SystemModel} has type $\bar{L} = (1, 3, 5)$ since data chunk $A$ is stored in these three servers. The set of all task types $\bar{L}$ is denoted by $\mathcal{L}$ defined as follows:
\vspace{-1mm}
$$\bar{L} \in \mathcal{L} = \{ (m_1, m_2, m_3) \in \mathcal{M}^3 : m_1 < m_2 < m_3\},$$
where $m_1, m_2,$ and $m_3$ are the three local servers \footnote{The analysis is not sensitive to the number of local servers. The default number of local servers in Hadoop is three, so we choose three local servers, but this assumption can be ignored without any change in the analysis.}. A task of type $\bar{L} = (m_1, m_2, m_3)$ receives faster average service from its local servers than from servers that do not have the data chunk. The reason is that the server without the data chunk has to fetch data associated to a task of type $\bar{L}$ from any of its local servers. According to the distance between the two servers, this fetching of the data can cause different amounts of delay. This fact brings the different levels of data locality into account. Obviously, the closer the two servers, the shorter the delay. Hence, the communication cost through the network and switches between two servers in the same rack is less than that between two servers in the same super rack (but different racks), and the cost for both is on average less than that between two servers in different super racks. Generally speaking, we propose the $N$ levels of data locality as follows: \\
\textbf{Service Process:} The non-preemptive service (processing) time of a task of type $\bar{L} = (m_1, m_2, m_3) \in \mathcal{L}$ is a random variable with cumulative distribution function (CDF)
\begin{itemize}[leftmargin=*]
\item $F_1$ with mean $\frac{1}{\alpha_1}$ if the task receives service from any server in the set $\bar{L} = \{m_1, m_2, m_3\}$, and we say that the task is $1$-local to these servers.
\item $F_n$ with mean $\frac{1}{\alpha_n}$ if the task receives service from any server in the set $\bar{L}_n$, defined in the following, and we say that the task is $n$-local to these servers, for $n \in \{2, 3, \cdots, N\}$,
\end{itemize}
where $\alpha_1 > \alpha_2 > \cdots > \alpha_N$. \\
In the data center structure example in Figure \ref{SystemModel}, the set $\bar{L}_2$ is the set of all servers that do not have the data saved on their own disk, but data is stored in another server in the same rack; and the set $\bar{L}_3$ is the set of all servers that do not have the data saved on their own disk, but data is stored in another server in another rack, but in the same super rack, and so on. \\
Remark. Note that the service time is not necessarily assumed to be geometrically distributed and can be arbitrary as long as it satisfies the decreasing property of the means mentioned above. \\ 
\textbf{Arrival Process:} The number of arriving tasks of type $\bar{L}$ at the beginning of time slot $t$ is denoted by $A_{\bar{L}}(t)$, which are assumed to be temporarily i.i.d. with mean $\lambda_{\bar{L}}$. The total number of arriving tasks at each time slot is assumed to be bounded by a constant $C_A$ and is assumed to be zero with a positive probability. The set of all arrival rates for different types of tasks is denoted by the vector $\mathbold{\lambda} = (\lambda_{\bar{L}}: \bar{L} \in \mathcal{L})$.

\subsection{An Outer Bound of the Capacity Region}
\label{CapacityRegionSection}

The arrival rate of type $\bar{L}$ tasks can be decomposed to ${(\lambda_{\bar{L}, m}, m \in \mathcal{M})}$, where $\lambda_{\bar{L}, m}$ denotes the arrival rate of type $\bar{L}$ tasks that are processed by server $m$. Obviously, $\sum_{m \in \mathcal{M}} \lambda_{\bar{L}, m} = \lambda_{\bar{L}}$. A necessary condition for an arrival rate vector $\mathbold{\lambda}$ to be supportable is that the total $1$-local, $2$-local, $\cdots$, $N$-local load on each server be strictly less than one as the following inequality suggests:
\begin{equation}
\label{necessarycondition}
\sum_{\bar{L} : m \in \bar{L}} \frac{\lambda_{\bar{L}, m}}{\alpha_1} + \sum_{\bar{L} : m \in \bar{L}_2} \frac{\lambda_{\bar{L}, m}}{\alpha_2} + \cdots + \sum_{\bar{L} : m \in \bar{L}_N} \frac{\lambda_{\bar{L}, m}}{\alpha_N} < 1, \ \forall m \in \mathcal{M}.
\end{equation}
Given this necessary condition, an outer bound of the capacity region is given by the set of all arrival rate vectors $\mathbold{\lambda}$ with a decomposition satisfying \eqref{necessarycondition}.

\begin{figure*}[t]
\centering
\includegraphics[scale=0.55]{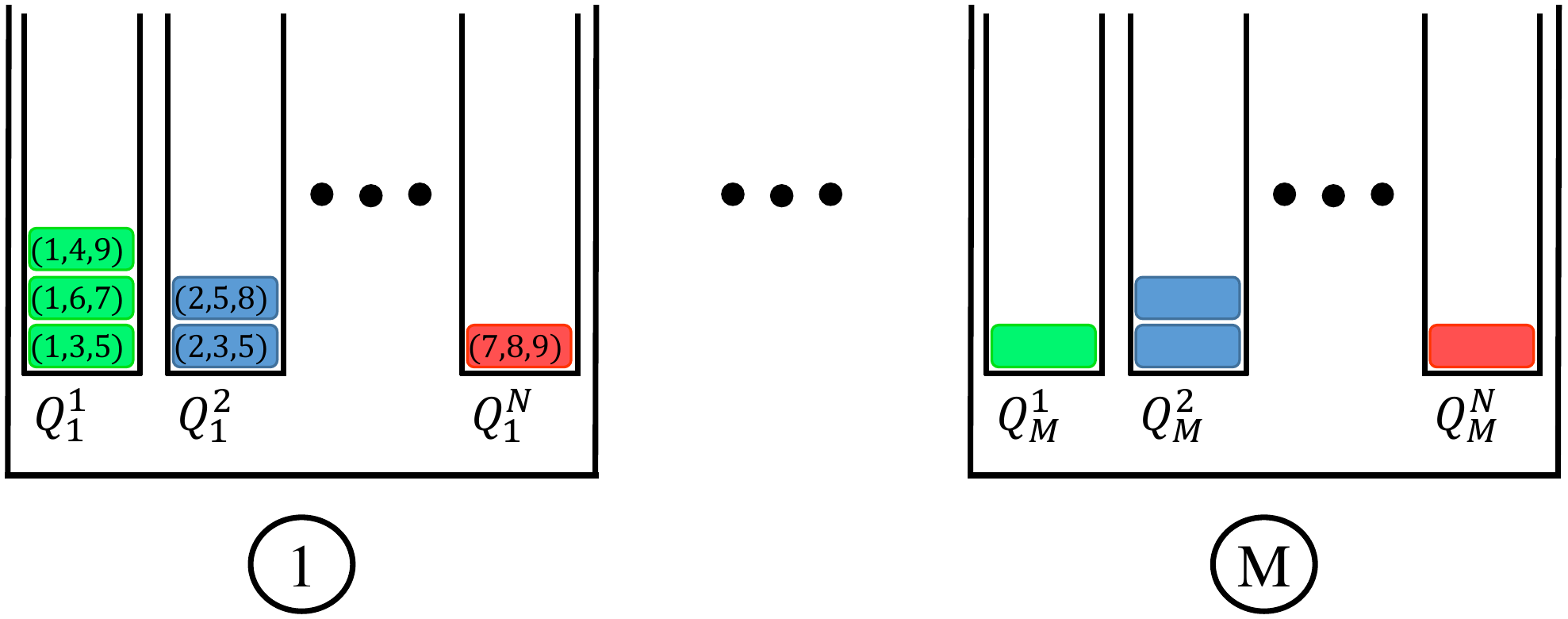}
\caption{The queueing structure when the GB-PANDAS algorithm is used.}
\label{QueueingStructure}
\end{figure*}

\vspace{-2.5mm}

\begin{equation}
\begin{aligned}
\label{capacityregion}
\Lambda = \big \{ & \boldsymbol{\lambda} = (\lambda_{\bar{L}} : \bar{L} \in \mathcal{L}) \ \big | \ \exists \lambda_{\bar{L}, m} \geq 0, \forall \bar{L} \in \mathcal{L}, \forall m \in \mathcal{M}, s.t. \\
& \lambda_{\bar{L}} = \sum_{m = 1}^{M} \lambda_{\bar{L}, m}, \ \forall \bar{L} \in \mathcal{L}, \\
& \sum_{\bar{L}: m \in \bar{L}} \frac{\lambda_{\bar{L}, m}}{\alpha_1} + \sum_{\bar{L}: m \in \bar{L}_2} \frac{\lambda_{\bar{L}, m}}{\alpha_2} + \cdots + \sum_{\bar{L}: m \in \bar{L}_N} \frac{\lambda_{\bar{L}, m}}{\alpha_N} < 1, \forall m \big \}.
\end{aligned}
\end{equation}
It is clear that to find $\Lambda$, we should solve a linear programming optimization problem. We will show in Section \ref{ThroughputOptimalitySection} that GB-PANDAS stabilizes the system as long as the arrival rate vector $\mathbold{\lambda}$ is inside $\Lambda$, which means that this outer bound of the capacity region is the capacity region itself.
In the following, Lemma \ref{Ecapacityregion} proposes a set which is equivalent to that in \eqref{capacityregion} which will be used in the throughput optimality proof of GB-PANDAS.
\begin{lemma}
\label{Ecapacityregion}
The following set $\bar{\Lambda}$ is equivalent to $\Lambda$ defined in equation \eqref{capacityregion}:

\begin{equation}
\begin{aligned}
\label{ECRR}
\bar{\Lambda} = \big \{ & \boldsymbol{\lambda} = (\lambda_{\bar{L}} : \bar{L} \in \mathcal{L}) \big | \exists \lambda_{\bar{L}, n, m} \geq 0, \forall \bar{L} \in \mathcal{L}, \forall n \in \bar{L}, \forall m \in \mathcal{M}, s.t. \\
& \lambda_{\bar{L}} = \sum_{n: n \in \bar{L}} \sum_{m = 1}^{M} \lambda_{\bar{L}, n, m}, \ \forall \bar{L} \in \mathcal{L}, \\
& \sum_{\bar{L}: m \in \bar{L}} \sum_{n: n \in \bar{L}} \frac{\lambda_{\bar{L}, n, m}}{\alpha_1} + \sum_{\bar{L}: m \in \bar{L}_2} \sum_{n: n \in \bar{L}} \frac{\lambda_{\bar{L}, n, m}}{\alpha_2} \\
& \ \ \ \ \ \ \ \ \ \ \ \ \ \ \ \ \ \ \ \ \ \ \ \ \ \ \ \ \ \ \ \ \ + \cdots + \sum_{\bar{L}: m \in \bar{L}_N} \sum_{n: n \in \bar{L}} \frac{\lambda_{\bar{L}, n, m}}{\alpha_N} < 1, \forall m \big \},
\end{aligned}
\end{equation}
where $\lambda_{\bar{L}, n, m}$ denotes the arrival rate of type $\bar{L}$ tasks that are $1$-local to server $n$ and is processed by server $m$. $\{\lambda_{\bar{L}, n, m}: \bar{L} \in \mathcal{L}, n \in \bar{L}, \text{and } m \in \mathcal{M}\}$ is a decomposition of the set of arrival rates $\{\lambda_{\bar{L}, m}: \bar{L} \in \mathcal{L} \text{ and } m \in \mathcal{M}\}$, where $\lambda_{\bar{L}, m} = \sum_{n \in \mathcal{M}} \lambda_{\bar{L}, n, m}$.
\end{lemma}
\textbf{Proof:}
We show that $\bar{\Lambda} \subset \Lambda$ and $\Lambda \subset \bar{\Lambda}$, which results in the equality of these two sets.\
\begin{itemize}[leftmargin=*]
	\item $\bar{\Lambda} \subset \Lambda$: If $\lambda \in \bar{\Lambda}$, there exists a decomposition $\{\lambda_{\bar{L}, n, m}: \bar{L} \in \mathcal{L}, n \in \bar{L}, \text{and } m \in \mathcal{M}\}$ such that the load on each server is less than one under this decomposition. Defining $\lambda_{\bar{L}, m} \equiv$ $\sum_{n: n \in \bar{L}}$ $\lambda_{\bar{L}, n, m}$, the arrival rate decomposition $\{\lambda_{\bar{L}, m}: \bar{L} \in \mathcal{L} \text{ and } m \in \mathcal{M}\}$ obviously satisfies the conditions in the definition of the set $\Lambda$, so $\lambda \in \Lambda$ which means that $\bar{\Lambda} \subset \Lambda$.
	\item $\Lambda \subset \bar{\Lambda}$: If $\lambda \in \Lambda$, there exists a decomposition $\{\lambda_{\bar{L}, m}: \bar{L} \in \mathcal{L} \text{ and } m \in \mathcal{M}\}$ such that the load on each server is less than one under this decomposition. Defining $\lambda_{\bar{L}, n, m} \equiv \frac{\lambda_{\bar{L}, m}}{|\bar{L}|}$, the arrival rate decomposition $\{\lambda_{\bar{L}, n, m}: \bar{L} \in \mathcal{L}, n \in \bar{L}, \text{and } m \in \mathcal{M}\}$ obviously satisfies the conditions in the definition of the set $\bar{\Lambda}$, so $\lambda \in \bar{\Lambda}$ which means that $\Lambda \subset \bar{\Lambda}$.
\end{itemize}

\section{The GB-PANDAS Algorithm}
\label{BP}
The central scheduler keeps $N$ queues per server as shown in Figure \ref{QueueingStructure}. The $N$ queues of the $m$-th server are denoted by $Q_m^1, Q_m^2, \cdots, Q_m^N$. Tasks that are routed to server $m$ and are $n$-local to this server are queued at queue $Q_m^n$. The length of this queue, defined as the number of tasks queued in this queue, at time slot $t$, is shown by $Q_m^n(t)$. The central scheduler maintains the length of all queues at all time slots, which is denoted by vector $\mathbold{Q}(t) =$ $\big ( Q_1^1(t),$ $Q_1^2(t),$ $\cdots,$ $Q_1^N(t),$ $\cdots,$ $Q_M^1(t),$ $Q_M^2(t),$ $\cdots,$ $Q_M^N(t)$ $\big )$. In the following, the workload on a server is defined which will be used in the statement of the GB-PANDAS algorithm.


\textbf{Workload of Server $m$:} Under the GB-PANDAS algorithm, server $m$ only processes tasks that are queued in its $N$ queues, that is $Q_m^1, Q_m^2, \cdots, Q_m^N$. As the processing time of an $n$-local task follows a distribution with CDF $F_n$ and mean $\frac{1}{\alpha_n}$, the expected time needed for server $m$ to process all tasks queued in its queues at time slot $t$ is given as follows:

\[
W_m(t) = \frac{Q_m^1(t)}{\alpha_1} + \frac{Q_m^2(t)}{\alpha_2} + \cdots + \frac{Q_m^N(t)}{\alpha_N}.
\]
We name $W_m(t)$ the \textit{workload} on the $m$-th server.

A load balancing algorithm consists of two parts, routing and scheduling. The routing policy determines the queue at which a new incoming task is queued until it receives service from a server. When a server becomes idle and so is ready to process another task, the scheduling policy determines the task receiving service from the idle server. The routing and scheduling policies of the GB-PANDAS algorithm are as follows:

\begin{itemize}[leftmargin=*]
\item \textbf{GB-PANDAS Routing (Weighted-Workload Routing):}
The incoming task of type $\bar{L}$ is routed to the corresponding sub-queue of server $m^*$ with the minimum weighted workload as defined in the following (ties are broken randomly):

\[
m^* = \underset{m \in \mathcal{M}}{\arg \min} \bigg \{ \frac{W_m(t)}{\alpha_1} I_{\{ m \in \bar{L} \}}, \frac{W_m(t)}{\alpha_2} I_{\{ m \in \bar{L}_2 \}}, \cdots, \frac{W_m(t)}{\alpha_N} I_{\{ m \in \bar{L}_N \}} \bigg \}.
\]
If this task of type $\bar{L}$ is $1$-local, $2$-local, $\cdots$, $N$-local to server $m^*$, it is queued at $Q_{m^*}^1$, $Q_{m^*}^2$, $\cdots$, $Q_{m^*}^N$, respectively.

\item \textbf{GB-PANDAS Scheduling (Prioritized Scheduling):}
The idle server $m$ is only scheduled to process a task from its own queues, $Q_m^1$, $Q_m^2$, $\cdots$, $Q_m^N$. A task that is $n$-local to server $m$ has a higher priority than a task that is $(n+1)$-local to server $m$ (for $1\leq n \leq N-1$). Hence, the idle server $m$ keeps processing a task from $Q_m^1$ until there are no more tasks available at this queue, then continues processing tasks queued at $Q_m^2$, and so on.
\end{itemize}

\subsection{Queue Dynamics}
Denote the number of arriving tasks at $Q_m^n$ at time slot $t$ by $A_m^n(t)$, where these tasks are $n$-local to server $m$. Recall the notation $A_{\bar{L}, m}(t)$ for the number of tasks of type $\bar{L}$ that are scheduled to server $m$. Then, we have the following relation between $A_m^n(t)$ and $A_{\bar{L}, m}(t)$:

\begin{equation}
\begin{aligned}
& A_m^1(t) = \sum_{\bar{L}: m \in \bar{L}} A_{\bar{L}, m}(t), \\
& A_m^n(t) = \sum_{\bar{L}: m \in \bar{L}_n} A_{\bar{L}, m}(t), \ \text{ for } \ 2 \leq n \leq N,
\label{sub-queue-arrival}
\end{aligned}
\end{equation}
where $\bar{L}$ is the set of $1$-local servers and $\bar{L}_n$ for $2 \leq n \leq N$ is the set of $n$-local servers to a task of type $\bar{L}$.
The number of tasks that receive service from server $m$ at time slot $t$ and are $n$-local to the server is denoted by $S_m^n(t)$ which is the number of departures from $Q_m^n$ (as a reminder, the service time of a task that is $n$-local to a server has CDF $F_n$). Then, the queue dynamics for any $m \in \mathcal{M}$ is as follows:

\begin{equation}
\begin{aligned}
\label{queueevolution}
& Q_m^n(t + 1) = Q_m^n(t) + A_m^n(t) - S_m^n(t), \ \text{ for } \ 1 \leq n \leq N - 1, \\
& Q_m^N(t + 1) = Q_m^N(t) + A_m^N(t) - S_m^N(t) + U_m(t),
\end{aligned}
\end{equation}
where $U_m(t) = \max \big \{ 0, S_m^N(t) - A_m^N(t) - Q_m^N(t) \big \}$ is the unused service of server $m$.

Note that the set of queue lengths $\{ \mathbold{Q}(t), t \geq 0 \}$ do not form a Markov chain since not having the information about how long a server has been processing a task and what type that task is, leads to $\mathbold{Q}(t + 1) | \mathbold{Q}(t) \not\perp \mathbold{Q}(t - 1)$. Note that the processing time of a task has a general CDF, not necessarily geometric distribution with memory-less property, so we do need to consider two parameters about the status of servers in the system as follows to be able to define a Markov chain.
\begin{itemize}[leftmargin=*]
	\item Let $\Psi_m(t)$ be the number of time slots at the beginning of time slot $t$ that server $m$ has spent on the currently in-service task. Note that $\Psi_m(t)$ is set to zero when server $m$ is done processing a task. Then the first working status vector, $\mathbold{\Psi}(t)$, is defined as follows:

$$\mathbold{\Psi}(t) = \big ( \Psi_1(t), \Psi_2(t), \cdots, \Psi_M(t) \big ).$$
	\item The second working status vector is $\mathbold{f}(t) =$ $\big ( f_1(t),$ $f_2(t),$ $\cdots,$ $f_M(t) \big )$, where

\[
f_m(t) = \begin{cases}
			   -1 \ \ \ \ \text{if server $m$ is idle,} \\
               \ \ 1 \ \ \ \ \ \text{if server $m$ processes a 1-local task from $Q_m^1$,} \\
			   \ \ 2 \ \ \ \ \ \text{if server $m$ processes a 2-local task from $Q_m^2$,} \\
			   \ \ \vdots \\
			   \ \ N \ \ \ \ \text{if server $m$ processes an N-local task from $Q_m^N$.}
         \end{cases}
\]
\end{itemize}

Define $\eta_m(t)$ as the scheduling decision for server $m$ at time slot $t$. If server $m$ finishes the processing of an in-service task at time slot $t$, we have $f_m(t^-) = -1$ and the central scheduler makes the scheduling decision $\eta_m(t)$ for the idle server $m$. Note that $\eta_m(t) = f_m(t)$ as long as server $m$ is processing a task. Then, we define the following vector:

$$\mathbold{\eta}(t) = \big ( \eta_1(t), \eta_2(t), \cdots, \eta_M(t) \big ).$$

As mentioned, since the service times have a general distribution with arbitrary CDF but not necessarily geometrically distributed, the queueing process --- or even both the queueing and $\mathbold{\eta}(t)$ processes --- do not form a Markov chain (one reason is that the service time does not have the memory-less property). Therefore, we consider the Markov chain $\big \{ \mathbold{Z}(t) = \big (\mathbold{Q}(t), \mathbold{\eta}(t), \mathbold{\Psi}(t) \big ), t \geq 0 \big \}$ and show that it is irreducible and aperiodic. The state space of this Markov chain is $\mathcal{S} = \mathbb{N}^{NM} \times \{ 1, 2, \cdots, N \}^M \times \mathbb{N}^{M}$. Assume the initial state of the Markov chain to be $\mathbold{Z}(0) = \big \{ 0_{NM \times 1}, N_{M \times 1}, 0_{M \times 1} \big \}$. \\
Irreducible: Since the CDF of the service times, $F_n$ for $1 \leq n \leq N$, are increasing, there exists a positive integer $\tau$ such that $F_n(\tau) > 0$ for $1 \leq n \leq N$. Moreover, the probability of zero arrival tasks is positive. Hence, for any state of the system, $\mathbold{Z} = \big ( \mathbold{Q}, \mathbold{\eta}, \mathbold{\Psi} \big )$, the probability of the event that each job gets processed in $\tau$ time slots and no tasks arrive at the system in $\tau \sum_{m = 1}^M \sum_{n = 1}^N Q_m^n$ time slots is positive. As a result, the initial state is reachable from any state in the state space and $\big \{ \mathbold{Z}(t) \big \}$ is irreducible. \\
Aperiodic: Since the probability of zero arriving tasks is positive, there is a positive probability of transition from the initial state to itself. Then, given that $\big \{ \mathbold{Z}(t) \big \}$ is irreducible, it is also aperiodic.

\section{Throughput Optimality}
\label{ThroughputOptimalitySection}
\begin{theorem}
\label{TTOBP}
The GB-PANDAS algorithm stabilizes a system with $N$ levels of data locality as long as the arrival rate is strictly inside the capacity region, which means that the Generalized Balanced-Pandas algorithm is throughput optimal.
\end{theorem}

\textbf{Proof:}
The throughput optimality proof of the GB-PANDAS algorithm for a system with $N$ levels of data locality and a general service time distribution follows an extension of the Foster-Lyapunov theorem as stated below. \\
\textbf{Extended Verion of the Foster-Lyapunov Theorem (Theorem 3.3.8 in \cite{srikant2013communication}):} Consider an irreducible Markov chain $\{ Z(t) \}$, where $t \in \mathbb{N}$, with a state space $\mathcal{S}$. If there exists a function $V: \mathcal{S} \rightarrow \mathcal{R}^+$, a positive integer $T \geq 1$, and a finite set $\mathcal{P} \subseteq \mathcal{S}$ satisfying the following condition:

\begin{equation}
\mathbb{E} \left [ V(Z(t_0 + T)) - V(Z(t_0)) | Z(t_0) = z \right ] \leq - \theta \mathbb{I}_{\{ z \in \mathcal{P}^c \}} + C \mathbb{I}_{\{ z \in \mathcal{P} \}},
\label{extendedLyapunov}
\end{equation}
for some $\theta > 0$ and $C < \infty$, then the irreducible Markov chain $\{ Z(t) \}$ is positive recurrent.

Consider the Markov chain $\big \{ \mathbold{Z}(t) = \big (\mathbold{Q}(t), \mathbold{\eta}(t), \mathbold{\Psi}(t) \big ), t \geq 0 \big \}$. As long as the arrival rate vector is strictly inside the outer bound of the capacity region, $\mathbold{\lambda} \in \Lambda$, and under using the GB-PANDAS algorithm, if we can prove that this Markov chain is positive recurrent, the distribution of $Z(t)$ converges to its stationary distribution when $t \rightarrow \infty$, which results in the stability of the system, so the throughput optimality of the GB-PANDAS algorithm will be proved.

As shown before, the Markov chain $Z(t)$ is irreducible and aperiodic for any arrival rate vector strictly inside the outer bound of the capacity region, $\mathbold{\lambda} \in \Lambda$. Hence, if we can find a Lyapunov function $V(.)$ satisfying the drift condition in the extended version of the Foster-Lyapunov theorem when using the GB-PANDAS algorithm, the stability of the system under this algorithm is proved. Lemmas \ref{lemma123}, \ref{lemma1234}, and \ref{lemma12345} followed by our choice of the Lyapunov function presented afterwards complete the proof.

Since $\Lambda$ is an open set, for any $\mathbold{\lambda} \in \Lambda$ there exists $\delta > 0$ such that $\mathbold{\lambda}' = (1 + \delta) \mathbold{\lambda} \in \Lambda$ which means that $\mathbold{\lambda}'$ satisfies the conditions in \eqref{capacityregion} and specifically the inequality \eqref{necessarycondition}. Then we have the following for any $m \in \mathcal{M}$:

\begin{equation}
\label{lambdaprime}
\sum_{\bar{L} : m \in \bar{L}} \frac{\lambda_{\bar{L}, m}}{\alpha_1} + \sum_{\bar{L} : m \in \bar{L}_2} \frac{\lambda_{\bar{L}, m}}{\alpha_2} + \cdots + \sum_{\bar{L} : m \in \bar{L}_N} \frac{\lambda_{\bar{L}, m}}{\alpha_N} < \frac{1}{1 + \delta}.
\end{equation}
The load decomposition $\{ \lambda_{\bar{L}, m} \}$ can be interpreted as one possibility of assigning the arrival rates to the $M$ servers so that the system becomes stable. We then define the ideal workload on each server $m$ under the load decomposition $\{ \lambda_{\bar{L}, m} \}$ as follows:

\begin{equation}
\label{workloadm}
w_m = \sum_{\bar{L}: m \in \bar{L}} \frac{\lambda_{\bar{L}, m}}{\alpha_1} + \sum_{\bar{L}: m \in \bar{L}_2} \frac{\lambda_{\bar{L}, m}}{\alpha_2} + \cdots + \sum_{\bar{L}: m \in \bar{L}_N} \frac{\lambda_{\bar{L}, m}}{\alpha_N}, \ \forall m \in \mathcal{M}.
\end{equation}
Let $\mathbold{ w} = (w_1, w_2, \cdots, w_M )$, where Lemmas \ref{lemma1234} and \ref{lemma12345} use this ideal workload on servers as an intermediary to later prove the throughput optimality of the GB-PANDAS algorithm.

The dynamic of the workload on server $m$, $W_m(.)$, is as follows:

\[
\begin{aligned}
W_m(t+1) & = \frac{Q_m^1(t+1)}{\alpha_1} + \frac{Q_m^2(t+1)}{\alpha_2} + \cdots + \frac{Q_m^N(t+1)}{\alpha_N} \\
& \overset{(a)}{=} \frac{Q_m^1(t) + A_m^1(t) - S_m^1(t)}{\alpha_1} + \frac{Q_m^2(t) + A_m^2(t) - S_m^2(t)}{\alpha_2} \\
& \ \ \ \ \ \ \ \ \ \ \ \ \ \ \ \ \ \ \ \ \ \ + \cdots + \frac{Q_m^N(t) + A_m^N(t) - S_m^N(t) + U_m(t)}{\alpha_N} \\
& = W_m(t) + \bigg (\frac{A_m^1(t)}{\alpha_1} + \frac{A_m^2(t)}{\alpha_2} + \cdots + \frac{A_m^N(t)}{\alpha_N} \bigg ) \\
& \ \ \ \ \ \ \ \ \ \ \ \ \ \ \ \ \ - \bigg (\frac{S_m^1(t)}{\alpha_1} + \frac{S_m^2(t)}{\alpha_2} + \cdots + \frac{S_m^N(t)}{\alpha_N} \bigg ) + \frac{U_m(t)}{\alpha_N} \\
& \overset{(b)}{=} W_m(t) + A_m(t) - S_m(t) + \widetilde{U}_m(t),
\end{aligned}
\]
where $(a)$ follows from the queue dynamic in \eqref{queueevolution} and $(b)$ is true by the following definitions:

\begin{equation}
\begin{aligned}
& A_m(t) = \frac{A_m^1(t)}{\alpha_1} + \frac{A_m^2(t)}{\alpha_2} + \cdots + \frac{A_m^N(t)}{\alpha_N}, \ \forall m \in \mathcal{M}, \\
& S_m(t) = \frac{S_m^1(t)}{\alpha_1} + \frac{S_m^2(t)}{\alpha_2} + \cdots + \frac{S_m^N(t)}{\alpha_N}, \ \forall m \in \mathcal{M}, \\
& \widetilde{U}_m(t) = \frac{U_m(t)}{\alpha_N}, \ \forall m \in \mathcal{M}.
\label{pseudoparameters}
\end{aligned}
\end{equation}
$\mathbold{A} = (A_1, A_2, \cdots, A_M)$, $\mathbold{S} = (S_1, S_2, \cdots, S_M)$, and $\widetilde{\mathbold{U}} =$ $(\widetilde{U}_1, \widetilde{U}_2,$ $\cdots,$ $\widetilde{U}_M)$ are the pseudo task arrival, service and unused service processes, respectively.

Then the dynamic of the workload on servers denoted by $\mathbold{W} = (W_1, W_2, \cdots, W_M)$ is as follows:

\begin{equation}
\label{evolW}
\mathbold{W}(t + 1) = \mathbold{W}(t) + \mathbold{A}(t) - \mathbold{S}(t) + \widetilde{\mathbold{U}}(t).
\end{equation}
We are now ready to propose Lemmas \ref{lemma123}, \ref{lemma1234}, \ref{lemma12345}, and \ref{lemma123456}.

\begin{lemma}
\label{lemma123}
\begin{equation*}
\langle \mathbold{W}(t) , \widetilde{\mathbold{U}}(t) \rangle = 0, \ \forall t.
\end{equation*}
\end{lemma}

\begin{lemma}
\label{lemma1234}
Under the GB-PANDAS routing policy, for any arrival rate vector strictly inside the outer bound of the capacity region, $\mathbold{\lambda} \in \Lambda$, and the corresponding workload vector of servers $\mathbold{w}$ defined in \eqref{workloadm}, we have the following for any $t_0$:

\begin{equation*}
\mathbb{E} \Big [ \langle \mathbold{W}(t), \mathbold{A}(t) \rangle - \langle \mathbold{W}(t), \mathbold{w} \rangle \Big | Z(t_0) \Big ] \leq 0, \ \forall t \geq 0.
\end{equation*}
\end{lemma}

\begin{lemma}
\label{lemma12345}
Under the GB-PANDAS routing policy, for any arrival rate vector strictly inside the outer bound of the capacity region, $\mathbold{\lambda} \in \Lambda$, and the corresponding workload vector of servers $\mathbold{w}$ defined in \eqref{workloadm} there exists $T_0 > 0$ such that for any $T \geq T_0$ we have the following:

\begin{equation*}
\begin{aligned}
& \mathbb{E} \left [ \sum_{t = t_0}^{t_0 + T - 1} \Big ( \langle \mathbold{W}(t), \mathbold{w} \rangle - \langle \mathbold{W}(t), \mathbold{S}(t) \rangle \Big ) \Big | Z(t_0) \right ] \\
\leq & - \theta_0 T || {\mathbold{Q}}(t_0) ||_1 + c_0, \ \forall t_0 \geq 0,
\end{aligned}
\end{equation*}
where the constants $\theta_0, c_0 > 0$ are independent of $Z(t_0)$.
\end{lemma}

\begin{lemma}
\label{lemma123456}
Under the GB-PANDAS routing policy, for any arrival rate vector strictly inside the outer bound of the capacity region, $\mathbold{\lambda} \in \Lambda$, and any $\theta_1 \in (0, 1)$, there exists $T_1 > 0$ such that the following is true for any $T \geq T_1$:

\[
\mathbb{E} \Big [ ||\mathbold{\Psi}(t_0 + T)||_1 - ||\mathbold{\Psi}(t_0)||_1 \Big | Z(t_0) \Big ] \leq -\theta_1 ||\mathbold{\Psi}(t_0)||_1 + M T, \ \forall t_0 \geq 0,
\]
where $||.||_1$ is $L^1$-norm.
\end{lemma}

We choose the following Lyapunov function, $V:\mathcal{P} \rightarrow \mathcal{R}^+$:

\[
V(Z(t)) = || \mathbold{W}(t) ||^2 + ||\mathbold{\Psi}(t)||_1,
\]
where $||.||$ and $||.||_1$ are the $L^2$ and $L^1$-norm, respectively. Then,

\begin{equation}
\begin{aligned}
& \mathbb{E} \Big [ V(Z(t_0 + T)) - V(Z(t_0)) \Big | Z(t_0) \Big ] \\
= \ & \mathbb{E} \Big [ || \mathbold{W}(t_0 + T) ||^2 - || \mathbold{W}(t_0) ||^2 \Big | Z(t_0) \Big ] \\
& \ \ \ \ \ \ \ \ \ \ \ \ \ \ \ \ \ \ \ \ \ \ \ \ \ \ \ \ \ \ \ \ \ \ \ \ \ \ \ \ + \mathbb{E} \Big [ || \mathbold{\Psi}(t_0 + T) ||_1 - || \mathbold{\Psi}(t_0) ||_1 \Big | Z(t_0) \Big ] \\
\overset{(a)}{=} \ & \mathbb{E} \left [ \sum_{t = t_0}^{t_0 + T - 1} \Big ( || \mathbold{W}(t + 1) ||^2 - || \mathbold{W}(t) ||^2 \Big ) \Big | Z(t_0) \right ] \\
& \ \ \ \ \ \ \ \ \ \ \ \ \ \ \ \ \ \ \ \ \ \ \ \ \ \ \ \ \ \ \ \ \ \ \ \ \ \ \ \ \ \ + \mathbb{E} \Big [ || \mathbold{\Psi}(t_0 + T) ||_1 - || \mathbold{\Psi}(t) ||_1 \Big | Z(t_0) \Big ] \\
\overset{(b)}{=} \ & \mathbb{E} \Bigg [ \sum_{t = t_0}^{t_0 + T - 1} \Big ( || \mathbold{A}(t) - \mathbold{S}(t) +\widetilde{\mathbold{U}}(t)||^2 + 2 \langle \mathbold{W}(t), \mathbold{A}(t) - \mathbold{S}(t) \rangle \\
& \ + 2 \langle \mathbold{W}(t), \widetilde{\mathbold{U}}(t) \rangle \Big ) \Big | Z(t_0) \Bigg ] + \mathbb{E} \Big [ || \mathbold{\Psi}(t_0 + T) ||_1 - || \mathbold{\Psi}(t) ||_1 \Big | Z(t_0) \Big ] \\
\overset{(c)}{\leq} \ & 2 \mathbb{E} \left [ \sum_{t = t_0}^{t_0 + T - 1} \Big ( \langle \mathbold{W}(t), \mathbold{A}(t) - \mathbold{S}(t) \rangle \Big ) \Big | Z(t_0) \right ] \\
& \ \ \ \ \ \ \ \ \ \ \ \ \ \ \ \ \ \ \ \ \ \ \ \ \ \ \ \ \ \ \ \ \ \ \ + \mathbb{E} \Big [ || \mathbold{\Psi}(t_0 + T) ||_1 - || \mathbold{\Psi}(t) ||_1 \Big | Z(t_0) \Big ] + c_1 \\
\overset{(d)}{=} \ & 2 \mathbb{E} \left [ \sum_{t = t_0}^{t_0 + T - 1} \Big ( \langle \mathbold{W}(t), \mathbold{A}(t) \rangle - \langle \mathbold{W}(t), \mathbold{w} \rangle \Big ) \Big | Z(t_0) \right ] \\
& \ \ \ \ \ \ \ \ \ \ \ \ \ \ \ \ \ \ \ \ \ + 2 \mathbb{E} \left [ \sum_{t = t_0}^{t_0 + T - 1} \Big ( \langle \mathbold{W}(t), \mathbold{w} \rangle - \langle \mathbold{W}(t), \mathbold{S}(t) \rangle \Big ) \Big | Z(t_0) \right ] \\
& \ \ \ \ \ \ \ \ \ \ \ \ \ \ \ \ \ \ \ \ \ \ \ \ \ \ \ \ \ \ \ \ \ + \mathbb{E} \Big [ || \mathbold{\Psi}(t_0 + T) ||_1 - || \mathbold{\Psi}(t) ||_1 \Big | Z(t_0) \Big ] + c_1,
\label{driftlyapunov}
\end{aligned}
\end{equation}
where $(a)$ is true by the telescoping property, $(b)$ follows by the dynamic of $\mathbold{W}(.)$ derived in \eqref{evolW}, $(c)$ follows by Lemma \ref{lemma123} and the fact that the task arrival is assumed to be bounded and the service and unused service are also bounded as the number of servers are finite, so the pseudo arrival, service, and unused service are also bounded, and therefore there exists a constant $c_1$ such that $|| \mathbold{A}(t) - \mathbold{S}(t) +\widetilde{\mathbold{U}}(t)||^2 \leq \frac{c_1}{T}$, and $(d)$ follows by adding and subtracting the intermediary term $\langle \mathbold{W}(t), \mathbold{w} \rangle$. \\
By choosing $T \geq \max\{T_0, T_1, \frac{\theta_1}{2\theta_0}\}$ and using Lemmas \ref{lemma1234}, \ref{lemma12345}, and \ref{lemma123456}, the drift of the Lyapunov function in \eqref{driftlyapunov} is the following:

\[
\begin{aligned}
& \mathbb{E} \Big [ V(Z(t_0 + T)) - V(Z(t_0)) \Big | Z(t_0) \Big ] \\
\leq & -\theta_1 \Big ( ||\mathbold{Q}(t_0)||_1 + ||\mathbold{\Psi}(t_0)||_1 \Big ) + c_2, \ \forall t_0,
\end{aligned}
\]
where $c_2 = 2c_0 + c_1 + MT$.\\
By choosing any positive constant $\theta_2 > 0$ let $\mathcal{P} = \big \{ Z = \big (\mathbold{Q}, \mathbold{\eta}, \mathbold{\Psi} \big ) \in \mathcal{S} : ||\mathbold{Q}||_1 + ||\mathbold{\Psi}||_1 \leq \frac{\theta_2 + c}{\theta_1} \big \}$, where $\mathcal{P}$ is a finite set of the state space. By this choice of $\mathcal{P}$, the condition \eqref{extendedLyapunov} in the extended version of the Foster-Lyapunov theorem holds by choices of $\theta = \theta_1$ and $C = c_2$, so the positive recurrence proof of the Markov chain and the throughput optimality proof of the GB-PANDAS algorithm are completed. Note that a corollary of this result is that $\Lambda$ is the capacity region of the system.

Note that in the proof of throughput optimality, we do not rely on the fact of using prioritized scheduling.
Therefore, for the purpose of throughput optimality, an idle server can serve any task in its $N$ sub-queues as $1$-local, $2$-local, $\cdots$, and $N$-local tasks decrease the expected workload at the same rate.
The prioritized scheduling is to minimize the mean task completion time experienced by tasks, which will be of interest in heavy-traffic optimality.
If fairness among jobs is of interest, we can assume sub-queues associated to jobs in each server and schedule an idle server to serve a task of the job which has the highest priority in terms of fairness. This does not affect the stability of the system.

\section{Simulation Results}
\label{SimulationResultsSection}
\begin{figure}[t]
\centering
\includegraphics[scale=0.2]{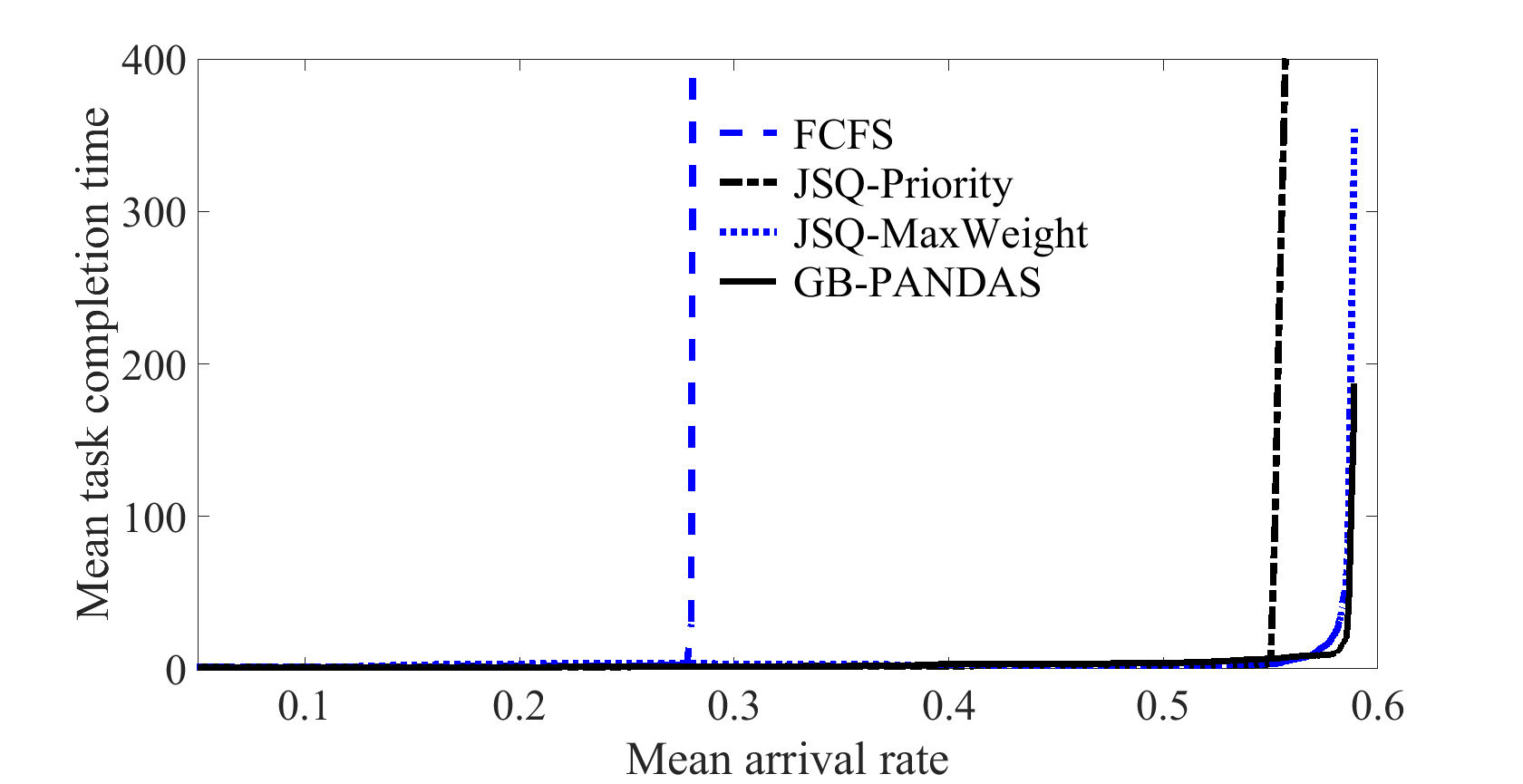}
\caption{Capacity region comparison of the algorithms.}
\label{3}
\end{figure}

\begin{figure}[t]
\centering
\includegraphics[scale=0.2]{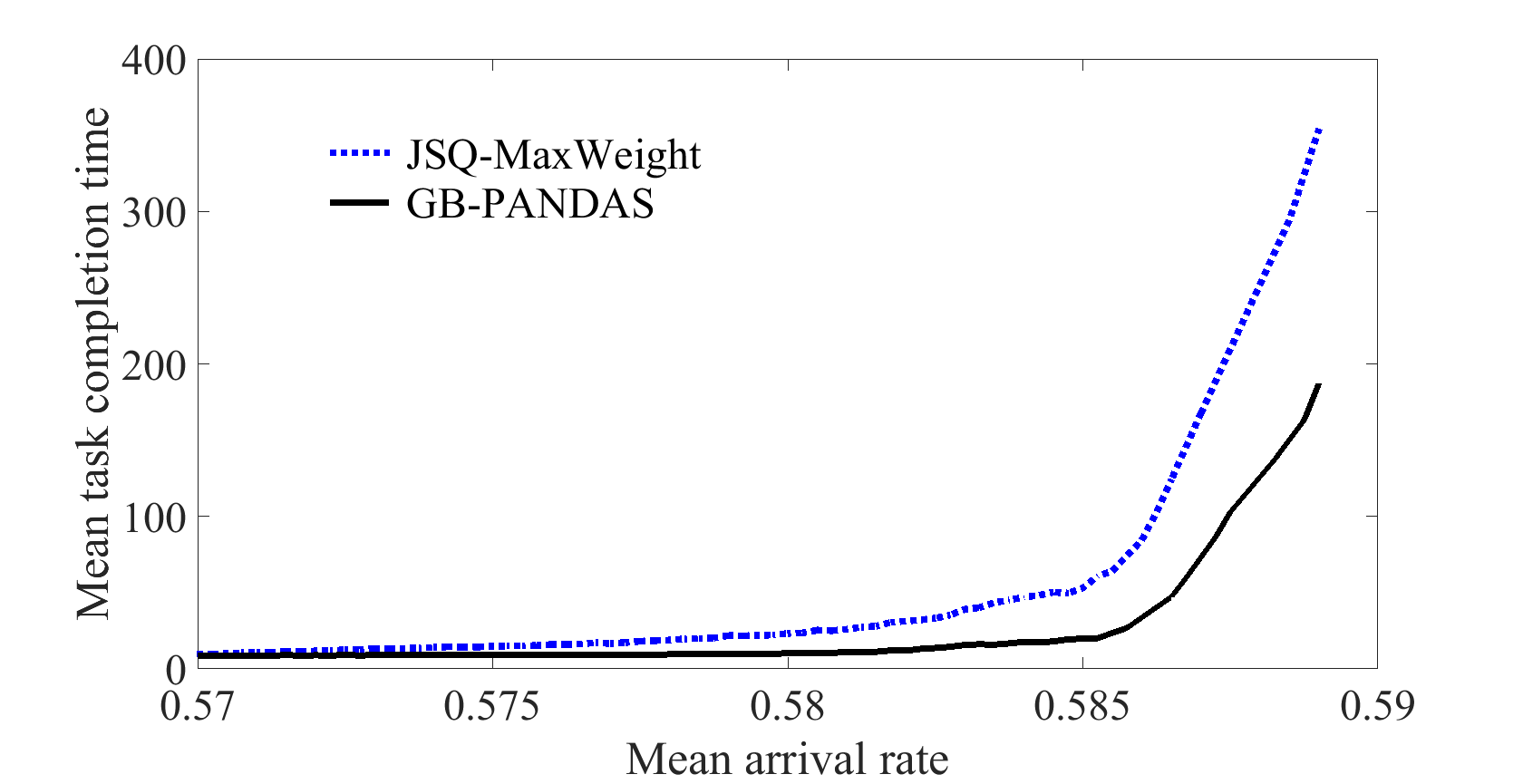}
\caption{Heavy-traffic performance.}
\label{33}
\end{figure}

In this section, we compare the simulated performance of our proposed algorithm, GB-PANDAS, against those of Hadoop's default FCFS scheduler, Join-the-Shortest-Queue-Priority (JSQ-Priority), and JSQ-MaxWeight algorithms. Consider a computing cluster with 5000 servers where each rack consists of 50 servers and each super rack includes 10 of the racks (so four levels of locality exist). We considered geometric and log-normal distributions for processing times and under both assumptions our algorithm outperforms others. Due to the space limit we only present the results for log-normal distribution (see \cite{reference} for more results). We assumed the i-local service time follows log-normal distribution with both mean and standard deviation equal to $\mu_i$ for $1 \leq i \leq 4$, where $\mu_1 = 1, \mu_2 = \frac{10}{9}, \mu_3 = \frac{5}{3},$ and $\mu_4 = 4$ (remote service is on average slower than local service by a factor of two to six times in data centers \cite{zaharia2010delay}, and we have chosen four times slowdown in our simulations). Figure \ref{3} shows the throughput performance of the four algorithms, where the y-axis shows the mean task completion time and the x-axis shows the mean arrival rate, i.e. $\frac{\sum_{\bar{L}} \lambda_{\bar{L}}}{M}$ (see lines $114-129$ in \href{https://www.dropbox.com/home/proof-general-distr/reference/3\%20-\%20generalized\%20-\%20large\%20variance/n_FCFS4?preview=PropA_skew.cc}{LINK} for details on the load). The GB-PANDAS and JSQ-MaxWeight algorithms are throughput optimal while FCFS and JSQ-Priority algorithms are not (note that JSQ-Priority is proven to be delay optimal for two locality levels, but it is not even throughput optimal for more locality levels). Figure \ref{33} compares the performance of the GB-PANDAS and JSQ-MaxWeight at high loads, where the first algorithm outperforms the latter by twofold. This significant improvement over JSQ-MaxWeight algorithm shows that JSQ-MaxWeight is not delay optimal and supports the possibility that the GB-PANDAS algorithm is delay optimal in a larger region than the JSQ-MaxWeight algorithm.

\begin{figure}[t]
\centering
\includegraphics[scale=0.2]{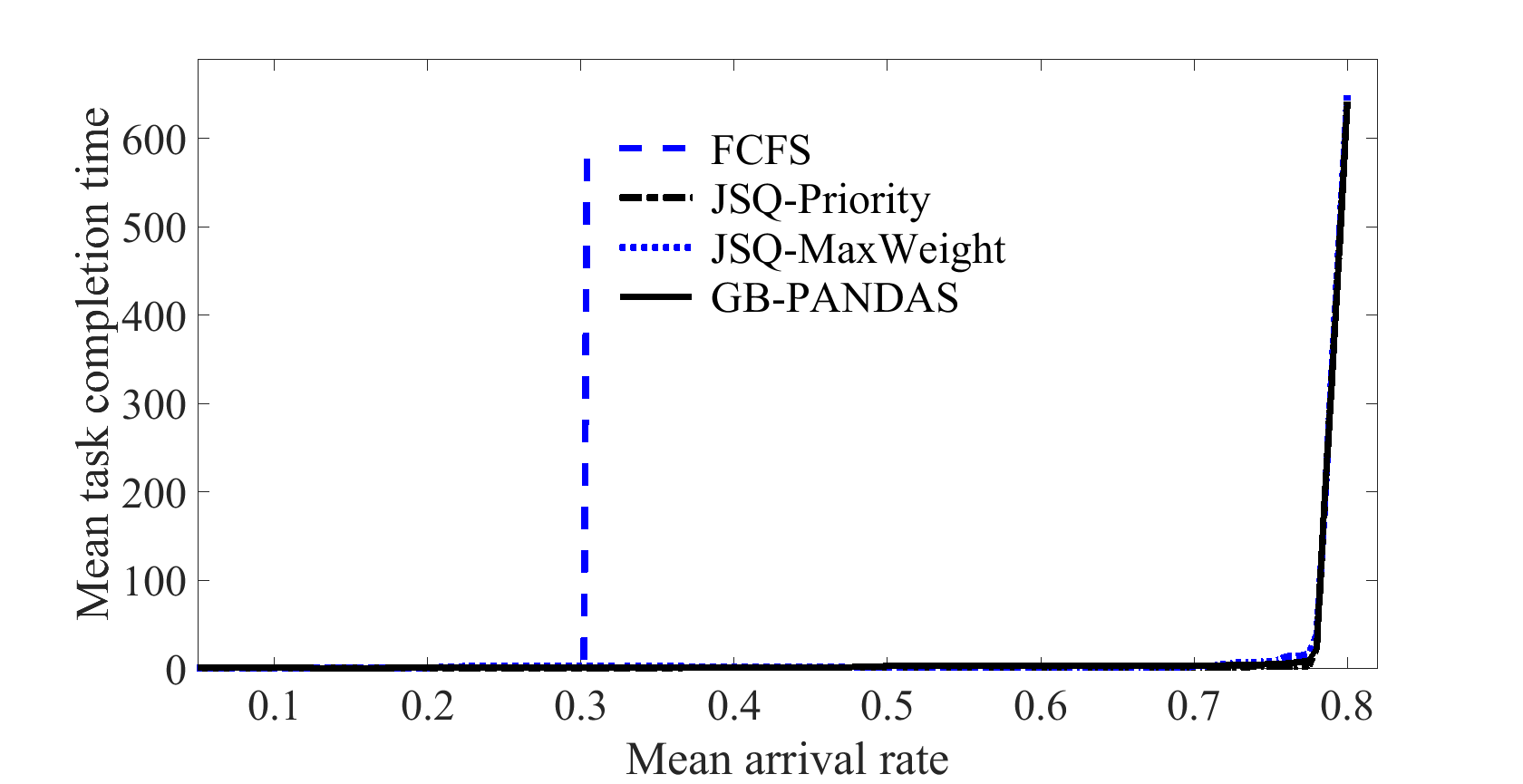}
\caption{Mean task completion time under a specific load.}
\label{4}
\end{figure}

By the intuition we got from the delay optimality proof of the JSQ-MaxWeight algorithm for two locality levels in \cite{wang2016maptask}, \cite{xie2016scheduling}, \cite{yekkehkhany2017near}, and \cite{xie2016schedulingPhD}, we simulated the system under a load for which we believe JSQ-MaxWeight is delay optimal. Figure \ref{4} shows the result for this specific load and we see that both the GB-PANDAS and JSQ-MaxWeight algorithms have the same performance at high loads, which again supports our guess on delay optimality of our proposed algorithm.

Note that Wang et al. \cite{wang2016maptask} showed that the JSQ-MaxWeight algorithm outperforms the Hadoop Fair Scheduler (HFS). Since our proposed algorithm outperforms JSQ-MaxWeight, we did not bring the HFS algorithm into our simulations.


\section{Related Work}
\label{RelatedWorkSection}
There is a huge body of work on scheduling for data centers considering data locality, which falls into two main categories: 1- Heuristic scheduling algorithms that do not even theoretically guarantee throughput optimality, see e.g. \cite{apache, isard2009quincy, zaharia2010delay, jin2011bar, he2011matchmaking, ibrahim2012maestro, polo2011resource, zaharia2008improving, noormohammadpour2017dccast, noormohammadpour2016dcroute}. Among these algorithms, the Fair Scheduler is the de facto standard in Hadoop \cite{zaharia2010delay}, but simple facts like the optimum delay time are not investigated for this algorithm (in fact the optimum delay varies in different loads). Other than map task scheduling for map-intensive jobs, heuristic algorithms like \cite{chen2012joint, tan2013coupling, lin2013joint} study the joint scheduling of map and reduce tasks. 2- Algorithms that theoretically guarantee throughput or delay optimality or both \cite{harrison1998heavy, harrison1999heavy, bell2001dynamic, bell2005dynamic, mandelbaum2004scheduling, stolyar2004maxweight, lu2017optimal, wang2016maptask, xie2015priority, xie2016scheduling}. The works by Harrison \cite{harrison1998heavy}, Harrison and Lopez \cite{harrison1999heavy}, and Bell and Williams \cite{bell2001dynamic, bell2005dynamic} on affinity scheduling not only require the knowledge of mean arrival rates of all task types, but also consider one queue per task type. In the data center example, if a task is replicated on three servers, the number of task types can be in the cubic order of number of servers, which causes unnecessary and intolerable complexity to the system. The MaxWeight algorithm (the generalized c$\mu$-rule) by Stolyar and Mandelbaum \cite{mandelbaum2004scheduling, stolyar2004maxweight} does not require the arrival rates, but still needs one queue per task type. The JSQ-MaxWeight algorithm by Wang et al. \cite{wang2016maptask} solves the per-task-type problem for a system with two levels of data locality. JSQ-MaxWeight is throughput optimal, but it is delay optimal for a special traffic scenario. The priority algorithm for near data scheduling \cite{xie2015priority} is both throughput and heavy-traffic optimal for systems with two locality levels. The weighted-workload routing and priority scheduling algorithm \cite{xie2016scheduling} for systems with three locality levels is shown to be throughput optimal and delay optimal (delay optimality needs the assumption $\alpha_2^2 > \alpha_1 \alpha_3$). In all of these studies, except \cite{wang2016maptask}, geometric distribution is assumed for service times.

\section{Conclusion and Future Work}
\label{ConclutionSection}
We proposed the GB-PANDAS algorithm for the affinity scheduling problem with a new queueing structure and proved its throughput optimality. The map task scheduling of map-intensive jobs in the MapReduce framework is an application of affinity scheduling, and our proposed algorithm is shown to have superior performance in simulation in this context. Investigating the delay optimality of our proposed algorithm in the region where JSQ-MaxWeight is delay optimal can be of interest. Furthermore, the existing delay optimality results of JSQ-MaxWeight \cite{wang2016maptask}, JSQ-Priority \cite{xie2015priority}, and weighted workload algorithm \cite{xie2016scheduling} are for exponentially distributed service times. It is interesting to investigate their delay optimality results under general distribution for service times. Considering the messaging costs, combination of power-of-d \cite{mukherjee2016universality} or join-idle-queue \cite{lu2011join} algorithms with GB-PANDAS may improve the performance.

\bibliographystyle{ACM-Reference-Format}
\bibliography{sigproc} 

\appendix
\section{Appendix}

\label{PROOFTTOBP}

\subsection{Proof of Lemma \ref{lemma123}}
\label{prooflemma123}
Lemma \ref{lemma123}:
\begin{equation*}
\langle \mathbold{W}(t) , \widetilde{\mathbold{U}}(t) \rangle = 0, \ \forall t.
\end{equation*}

\textbf{Proof:}
The expression simplifies as follows:
\[
\langle \mathbold{W}(t) , \widetilde{\mathbold{U}}(t) \rangle = \sum_m \bigg ( \frac{Q_m^1(t)}{\alpha_1} + \frac{Q_m^2(t)}{\alpha_2} + \cdots + \frac{Q_m^N(t)}{\alpha_N} \bigg ) \frac{U_m(t)}{\alpha_N}.
\]
Note that for any server $m$, $U_m(t)$ is either zero or positive. For the first case it is obvious that $\big ( \frac{Q_m^1(t)}{\alpha_1} + \frac{Q_m^2(t)}{\alpha_2} + \cdots + \frac{Q_m^N(t)}{\alpha_N} \big ) \frac{U_m(t)}{\alpha_N} = 0 $.
In the latter case where $U_m(t) > 0$, all sub-queues of server $m$ are empty which again results in $\big ( \frac{Q_m^1(t)}{\alpha_1} + \frac{Q_m^2(t)}{\alpha_2} + \cdots + \frac{Q_m^N(t)}{\alpha_N} \big ) \frac{U_m(t)}{\alpha_N} = 0 $.
Therefore, $\langle \mathbold{W}(t) , \widetilde{\mathbold{U}}(t) \rangle = 0$ for all time slots.

\subsection{Proof of Lemma \ref{lemma1234}}
\label{prooflemma1234}
Lemma \ref{lemma1234}:
Under the GB-PANDAS algorithm, for any arrival rate vector strictly inside the outer bound of the capacity region, $\mathbold{\lambda} \in \Lambda$, and the corresponding workload vector of servers $\mathbold{w}$ defined in \eqref{workloadm}, we have the following for any $t_0$:
\begin{equation*}
\mathbb{E} \Big [ \langle \mathbold{W}(t), \mathbold{A}(t) \rangle - \langle \mathbold{W}(t), \mathbold{w} \rangle \Big | Z(t_0) \Big ] \leq 0, \ \forall t \geq 0.
\end{equation*}

\textbf{Proof:}
The minimum weighted workload for type $\bar{L}$ task, where $\bar{L} \in \mathcal{L}$, at time slot $t$ is defined as follows:
\begin{equation*}
W_{\bar{L}}^*(t) = \min_{m \in \mathcal{M}} \bigg \{ \frac{W_m(t)}{\alpha_1} I_{\{ m \in \bar{L} \}}, \frac{W_m(t)}{\alpha_2} I_{\{ m \in \bar{L}_2 \}}, \cdots, \frac{W_m(t)}{\alpha_N} I_{\{ m \in \bar{L}_N \}} \bigg \}.
\end{equation*}
According to the routing policy of the GB-PANDAS algorithm, an incoming task of type $\bar{L}$ at the beginning of time slot $t$ is routed to the corresponding sub-queue of server $m^*$ with the minimum weighted workload $W_{\bar{L}}^*$. Therefore, for any type $\bar{L}$ task we have the following:
\begin{equation}
\label{MWW}
\begin{aligned}
& \frac{W_m(t)}{\alpha_1} \geq W_{\bar{L}}^*(t), \ \forall m \in \bar{L}, \\
& \frac{W_m(t)}{\alpha_n} \geq W_{\bar{L}}^*(t), \ \forall m \in \bar{L}_n, \ \text{ for } 2 \leq n \leq N.
\end{aligned}
\end{equation}

In other words, a type $\bar{L}$ task does not join a server with a weighted workload greater than $W_{\bar{L}}^*$.
Using the fact that $\mathbold{W}(t)$ and $\mathbold{A}(t)$ are conditionally independent of $Z(t_0)$ given $Z(t)$, and also following the definitions of pseudo task arrival process $\mathbold{A}(t)$ in \eqref{pseudoparameters} and the arrival of an $n$-local type task to the $m$-th server $A_m^n(t)$ in \eqref{sub-queue-arrival}, we have the following:
\begin{equation}
\begin{aligned}
\label{eq1}
& \mathbb{E} \big [ \langle \mathbold{W}(t), \mathbold{A}(t) \rangle | Z(t_0) \big ] = \mathbb{E} \Big [ \mathbb{E} \big [ \langle \mathbold{W}(t), \mathbold{A}(t) \rangle | Z(t) \big ] \Big | Z(t_0) \Big ] \\
& \hspace{-1mm} = \mathbb{E} \left [ \mathbb{E} \left [ \sum_m W_m(t) \left ( \frac{A_m^1(t)}{\alpha_1} + \frac{A_m^2(t)}{\alpha_2} + \cdots + \frac{A_m^N(t)}{\alpha_N} \right ) \bigg | Z(t) \right ] \bigg | Z(t_0) \right ] \\
& \hspace{-1mm} = \mathbb{E} \bigg [ \mathbb{E} \bigg [ \sum_m W_m(t) \bigg ( \frac{1}{\alpha_1} \sum_{\bar{L}: m \in \bar{L}} A_{\bar{L}, m}(t) + \frac{1}{\alpha_2} \sum_{\bar{L}: m \in \bar{L}_2} A_{\bar{L}, m}(t) \\
& \ \ \ \ \ \ \ \ \ \ \ \ \ \ \ \ \ \ \ \ \ \ \ \ \ \ \ \ \ \ \ \ \ \ \ \ \ \ + \cdots + \frac{1}{\alpha_N} \sum_{\bar{L}: m \in \bar{L}_N} A_{\bar{L}, m}(t) \bigg ) \bigg | Z(t) \bigg ] \bigg | Z(t_0) \bigg ] \\
& \hspace{-1mm} \overset{(a)}{=} \mathbb{E} \bigg [ \mathbb{E} \bigg [ \sum_{\bar{L} \in \mathcal{L}} \bigg ( \sum_{m: m \in \bar{L}} \frac{W_m(t)}{\alpha_1} A_{\bar{L}, m}(t) + \sum_{m: m \in \bar{L}_2} \frac{W_m(t)}{\alpha_2} A_{\bar{L}, m}(t)  \\ 
& \ \ \ \ \ \ \ \ \ \ \ \ \ \ \ \ \ \ \ \ \ \ \ \ \ \ \ \ \ \ \ \ \ + \cdots + \sum_{m: m \in \bar{L}_N} \frac{W_m(t)}{\alpha_N} A_{\bar{L}, m}(t) \bigg ) \bigg | Z(t) \bigg ] \bigg | Z(t_0) \bigg ] \\
& \overset{(b)}{=} \mathbb{E} \left [ \mathbb{E} \left [ \sum_{\bar{L} \in \mathcal{L}} W_{\bar{L}}^*(t) A_{\bar{L}}(t) \Big | Z(t) \right ] \bigg | Z(t_0) \right ] = \sum_{\bar{L} \in \mathcal{L}} W_{\bar{L}}^*(t) \lambda_{\bar{L}},
\end{aligned}
\end{equation}

where $(a)$ is true by changing the order of the summations, and $(b)$ follows by the GB-PANDAS routing policy which routes type $\bar{L}$ task to the server with the minimum weighted workload, $W_{\bar{L}}^*$.
Furthermore, using the definition of the ideal workload on a server in \eqref{workloadm} we have the following:
\begin{equation}
\begin{aligned}
\label{eq2}
& \mathbb{E} \big [ \langle \mathbold{W}(t), \mathbold{w} \rangle | Z(t) \big ] = \sum_{m = 1}^M W_m(t) w_m \\
= \ & \sum_m W_m(t) \bigg ( \sum_{\bar{L}: m \in \bar{L}} \frac{\lambda_{\bar{L}, m}}{\alpha_1} + \sum_{\bar{L}: m \in \bar{L}_2} \frac{\lambda_{\bar{L}, m}}{\alpha_2} + \cdots + \sum_{\bar{L}: m \in \bar{L}_N} \frac{\lambda_{\bar{L}, m}}{\alpha_N} \bigg ) \\
\overset{(a)}{=} & \sum_{\bar{L} \in \mathcal{L}} \bigg ( \sum_{m: m \in \bar{L}} \frac{W_m(t)}{\alpha_1} \lambda_{\bar{L}, m} + \sum_{m: m \in \bar{L}_2} \frac{W_m(t)}{\alpha_2} \lambda_{\bar{L}, m} \\
& \ \ \ \ \ \ \ \ \ \ \ \ \ \ \ \ \ \ \ \ \ \ \ \ \ \ \ \ \ \ \ \ \ \ \ \ \ \ \ \ \ \ \ \ \ \ \ \ \ \ \ \ \ + \cdots + \sum_{m: m \in \bar{L}_N} \frac{W_m(t)}{\alpha_N} \lambda_{\bar{L}, m} \bigg ) \\
\overset{(b)}{\geq} & \sum_{\bar{L} \in \mathcal{L}} \sum_{m \in \mathcal{M}} W_{\bar{L}}^*(t)\lambda_{\bar{L}, m} = \sum_{\bar{L} \in \mathcal{L}} W_{\bar{L}}^*(t) \lambda_{\bar{L}},
\end{aligned}
\end{equation}
where $(a)$ is true by changing the order of summations, and $(b)$ follows from \eqref{MWW}.
Lemma \ref{lemma1234} is concluded from equations \eqref{eq1} and \eqref{eq2}.


\subsection{Proof of Lemma \ref{lemma12345}}
\label{prooflemma12345}
Lemma \ref{lemma12345}:
Under the GB-PANDAS algorithm, for any arrival rate vector strictly inside the outer bound of the capacity region, $\mathbold{\lambda} \in \Lambda$, and the corresponding workload vector of servers $\mathbold{w}$ defined in \eqref{workloadm} there exists $T_0 > 0$ such that for any $T \geq T_0$ we have the following:

\begin{equation*}
\begin{aligned}
& \mathbb{E} \left [ \sum_{t = t_0}^{t_0 + T - 1} \Big ( \langle \mathbold{W}(t), \mathbold{w} \rangle - \langle \mathbold{W}(t), \mathbold{S}(t) \rangle \Big ) \Big | Z(t_0) \right ] \\
\leq & - \theta_0 T || {\mathbold{Q}}(t_0) ||_1 + c_0, \ \forall t_0 \geq 0,
\end{aligned}
\end{equation*}
where the constants $\theta_0, c_0 > 0$ are independent of $Z(t_0)$.

\textbf{Proof:}
By our assumption on boundedness of arrival and service processes, there exists a constant $C_A$ such that for any $t_0, t,$ and $T$ with $t_0 \leq t \leq t_0 + T$, we have the following:

\begin{equation}
W_m(t_0) - \frac{T}{\alpha_N} \leq W_m(t) \leq W_m(t_0) + \frac{T C_A}{\alpha_N}, \ \forall m \in \mathcal{M}.
\label{boundT}
\end{equation}
On the other hand, by \eqref{lambdaprime} the ideal workload on a server defined in \eqref{workloadm} can be bounded as follows:

\begin{equation}
w_m \leq \frac{1}{1 + \delta}, \ \forall m \in \mathcal{M}.
\label{workloadbound}
\end{equation}
Hence,

\begin{equation}
\begin{aligned}
& \mathbb{E} \left [ \sum_{t = t_0}^{t_0 + T - 1} \Big ( \langle \mathbold{W}(t), \mathbold{w} \rangle \Big ) \Big | Z(t_0) \right ]
 \\
= \ & \mathbb{E} \left [ \sum_{t = t_0}^{t_0 + T - 1} \left ( \sum_{m = 1}^M W_m(t) w_m \right ) \bigg | Z(t_0) \right ] \\
\overset{(a)}{\leq} & \ T \sum_{m = 1}^M \Big ( W_m(t_0) w_m \Big ) + \frac{MT^2C_A}{\alpha_N} \\
\overset{(b)}{\leq} & \ \frac{T}{1 + \delta} \sum_m W_m(t_0) + \frac{MT^2C_A}{\alpha_N},
\label{eq21}
\end{aligned}
\end{equation}
where $(a)$ is true by bringing the inner summation on $m$ out of the expectation and using the boundedness property of the workload in equation \eqref{boundT}, and $(b)$ is true by equation \eqref{workloadbound}.

Before investigating the second term, $\mathbb{E} \Big [$ $\sum_{t = t_0}^{t_0 + T - 1}$ $\Big ( \langle \mathbold{W}(t), \mathbold{S}(t) \rangle \Big )$ $\Big | Z(t_0)  \Big ]$, we propose the following lemma which will be used in lower bounding this second term.

\begin{lemma}
For any server $m \in \mathcal{M}$ and any $t_0$, we have the following:
$$\lim_{T \rightarrow \infty} \frac{ \mathbb{E} \left [ \sum_{t = t_0}^{t_0 + T - 1} \bigg ( \frac{S_m^1(t)}{\alpha_1} + \frac{S_m^2(t)}{\alpha_2} + \cdots + \frac{S_m^N(t)}{\alpha_N} \bigg ) \bigg | Z(t_0) \right ]}{T} = 1.$$
\label{asymptoticT}
\end{lemma}

We then have the following:

\begin{equation}
\label{eq22}
\begin{aligned}
& \mathbb{E} \left [ \sum_{t = t_0}^{t_0 + T - 1} \Big ( \langle \mathbold{W}(t), \mathbold{S}(t) \rangle \Big ) \Big | Z(t_0)  \right ] \\
= \ & \mathbb{E} \Bigg [ \sum_{t = t_0}^{t_0 + T - 1} \sum_{m = 1}^{M} \Bigg ( W_m(t) \bigg ( \frac{S_m^1(t)}{\alpha_1} + \frac{S_m^2(t)}{\alpha_2} \\
& \ \ \ \ \ \ \ \ \ \ \ \ \ \ \ \ \ \ \ \ \ \ \ \ \ \ \ \ \ \ \ \ \ \ \ \ \ \ \ \ \ \ \ \ \ \ \ \ \ \ \ \ \ \ \ \ \ \ \ + \cdots + \frac{S_m^N(t)}{\alpha_N} \bigg ) \Bigg ) \bigg | Z(t_0) \Bigg ] \\
\overset{(a)}{ \geq } & \ \sum_{m = 1}^{M} \Bigg ( W_m(t_0) \mathbb{E} \Bigg [ \sum_{t = t_0}^{t_0 + T - 1} \bigg ( \frac{S_m^1(t)}{\alpha_1} + \frac{S_m^2(t)}{\alpha_2} \\
& \ \ \ \ \ \ \ \ \ \ \ \ \ \ \ \ \ \ \ \ \ \ \ \ \ \ \ \ \ \ \ \ \ \ \ \ \ \ \ \ \ \ \ \ \ \ \ \ \ \ \ \ \ \ \ \ \ \ \ + \cdots + \frac{S_m^N(t)}{\alpha_N} \bigg ) \bigg | Z(t_0) \Bigg ] \Bigg ) \\
& - \frac{T}{\alpha_N} \sum_{m = 1}^{M} \mathbb{E} \left [ \sum_{t = t_0}^{t_0 + T - 1} \bigg ( \frac{S_m^1(t)}{\alpha_1} + \frac{S_m^2(t)}{\alpha_2} + \cdots + \frac{S_m^N(t)}{\alpha_N} \bigg ) \bigg | Z(t_0) \right ], \\
\end{aligned}
\end{equation}
where $(a)$ follows by bringing the inner summation on $m$ out of the expectation and using the boundedness property of the workload in equation \eqref{boundT}.

Using Lemma \ref{asymptoticT}, for any $0 < \epsilon_0 < \frac{\delta}{1 + \delta}$, there exists $T_0$ such that for any $T \geq T_0$, we have the following for any server $m \in \mathcal{M}$:

$$1 - \epsilon_0 \leq \frac{ \mathbb{E} \left [ \sum_{t = t_0}^{t_0 + T - 1} \bigg ( \frac{S_m^1(t)}{\alpha_1} + \frac{S_m^2(t)}{\alpha_2} + \cdots + \frac{S_m^N(t)}{\alpha_N} \bigg ) \bigg | Z(t_0) \right ]}{T} \leq 1 + \epsilon_0.$$
Then continuing on equation \eqref{eq22} we have the following:

\begin{equation}
\label{eq22222}
\begin{aligned}
& \mathbb{E} \left [ \sum_{t = t_0}^{t_0 + T - 1} \Big ( \langle \mathbold{W}(t), \mathbold{S}(t) \rangle \Big ) \Big | Z(t_0)  \right ] \\
\geq & \ T (1 - \epsilon_0)\sum_{m = 1}^{M} W_m(t_0) - \frac{MT^2(1 + \epsilon_0)}{\alpha_N}. \\
\end{aligned}
\end{equation}

Then Lemma \ref{lemma12345} is concluded as follows by using equations \eqref{eq21} and \eqref{eq22222} and picking $c_0 = \frac{M T^2}{\alpha_N} (C_A + 1 + \epsilon_0)$ and $\theta_0 = \frac{1}{\alpha_1} \left ( \frac{\delta}{1 + \delta} - \epsilon_0 \right )$, where by our choice of $\epsilon_0$ we have $\theta_0 > 0$:

\begin{equation*}
\begin{aligned}
& \mathbb{E} \left [ \sum_{t = t_0}^{t_0 + T - 1} \Big ( \langle \mathbold{W}(t), \mathbold{w} \rangle - \langle \mathbold{W}(t), \mathbold{S}(t) \rangle \Big ) \Big | Z(t_0) \right ] \\
\leq & - T \left ( \frac{\delta}{1 + \delta} - \epsilon_0 \right ) \sum_{m = 1}^{M} W_m(t_0) + \frac{M T^2}{\alpha_N} (C_A + 1 + \epsilon_0) \\
\overset{(a)}{\leq} & - \frac{T}{\alpha_1} \left ( \frac{\delta}{1 + \delta} - \epsilon_0 \right ) \sum_{m = 1}^{M} \Big (Q_m^1(t_0) + Q_m^2(t_0) + \cdots + Q_m^N(t_0) \Big ) + c_0 \\
\leq & - \theta_0 T || {\mathbold{Q}}(t_0) ||_1 + c_0, \ \forall t_0 \geq 0,
\end{aligned}
\end{equation*}
where $(a)$ is true as $W_m(t_0) \geq \frac{Q_m^1(t_0) + Q_m^2(t_0) + \cdots + Q_m^N(t_0)}{\alpha_1}$.

\subsection{Proof of Lemma \ref{asymptoticT}}
\label{proofasymptoticT}
Lemma \ref{asymptoticT}:
For any server $m \in \mathcal{M}$ and any $t_0$, we have the following:

$$\lim_{T \rightarrow \infty} \frac{ \mathbb{E} \left [ \sum_{t = t_0}^{t_0 + T - 1} \bigg ( \frac{S_m^1(t)}{\alpha_1} + \frac{S_m^2(t)}{\alpha_2} + \cdots + \frac{S_m^N(t)}{\alpha_N} \bigg ) \bigg | Z(t_0) \right ]}{T} = 1.$$

\textbf{Proof:}
Let $t_m^*$ be the first time slot after or at time slot $t_0$ at which server $m$ becomes idle, and so is available to serve another task; that is,

\begin{equation}
t_m^* = \min\{ \tau: \tau \geq t_0, \Psi_m(\tau) = 0 \},
\label{tm**}
\end{equation}
where, as a reminder, $\Psi_m(\tau)$ is the number of time slots that the $m$-th server has spent on the task that is receiving service from this server at time slot $\tau$. Note that the CDF of the service time distributions are given by $F_n, n \in \{1, 2, \cdots, N\}$ where they all have finite means $\alpha_n < \infty$; therefore, $t_m^* < \infty$. We then have the following by considering the bounded service:

\begin{equation}
\begin{aligned}
& \frac{ \mathbb{E} \left [ \sum_{t = t_m^*}^{t_m^* + T - 1} \bigg ( \frac{S_m^1(t)}{\alpha_1} + \frac{S_m^2(t)}{\alpha_2} + \cdots + \frac{S_m^N(t)}{\alpha_N} \bigg ) \bigg | Z(t_0) \right ] - \frac{t_m^* - t_0}{\alpha_N} + \frac{1}{\alpha_1}}{T} \\
& \leq \frac{ \mathbb{E} \left [ \sum_{t = t_0}^{t_0 + T - 1} \bigg ( \frac{S_m^1(t)}{\alpha_1} + \frac{S_m^2(t)}{\alpha_2} + \cdots + \frac{S_m^N(t)}{\alpha_N} \bigg ) \bigg | Z(t_0) \right ]}{T} \\
& \leq \frac{ \mathbb{E} \left [ \sum_{t = t_m^*}^{t_m^* + T - 1} \bigg ( \frac{S_m^1(t)}{\alpha_1} + \frac{S_m^2(t)}{\alpha_2} + \cdots + \frac{S_m^N(t)}{\alpha_N} \bigg ) \bigg | Z(t_0) \right ] + \frac{1}{\alpha_N}}{T},
\label{renewalProcess}
\end{aligned}
\end{equation}
where by boundedness of $t_m^*, \alpha_1,$ and $\alpha_N$, it is obvious that $\lim_{T \rightarrow \infty}$ $\frac{- \frac{t_m^* - t_0}{\alpha_N} + \frac{1}{\alpha_1}}{T} = 0$ and $\lim_{T \rightarrow \infty} \frac{\frac{1}{\alpha_N}}{T} = 0$. Hence, by taking the limit of the terms in equation \eqref{renewalProcess} as $T$ goes to infinity, we have the following:

\begin{equation}
\begin{aligned}
& \lim_{T \rightarrow \infty} \frac{ \mathbb{E} \left [ \sum_{t = t_0}^{t_0 + T - 1} \bigg ( \frac{S_m^1(t)}{\alpha_1} + \frac{S_m^2(t)}{\alpha_2} + \cdots + \frac{S_m^N(t)}{\alpha_N} \bigg ) \bigg | Z(t_0) \right ]}{T} \\
= & \lim_{T \rightarrow \infty} \frac{ \mathbb{E} \left [ \sum_{t = t_m^*}^{t_m^* + T - 1} \bigg ( \frac{S_m^1(t)}{\alpha_1} + \frac{S_m^2(t)}{\alpha_2} + \cdots + \frac{S_m^N(t)}{\alpha_N} \bigg ) \bigg | Z(t_0) \right ]}{T}.
\label{t_m^*}
\end{aligned}
\end{equation}
Considering the service process as a renewal process, given the scheduling decisions at the end of the renewal intervals in $[t_m^*, t_m^* + T - 1]$, all holding times for server $m$ to give service to tasks in its queues are independent. We elaborate on this in the following.

We define renewal processes, $N_m^n(t), \ n \in \{1, 2, \cdots, N\}$, as follows, where $t$ is an integer valued number:

Let $H_m^n(l)$ be the holding time (service time) of the $l$-th task that is $n$-local to server $m$ after time slot $t_m^*$ receiving service from serve $m$, and call $\{H_m^n(l), l \geq 1\}$ the holding process of $n$-local type task ($n \in \{1, 2, \cdots, N\}$). Then define $J_m^n(l) = \sum_{i = 1}^{l} H_m^n(l)$ for $l \geq 1$, and let $J_m^n(0) = 0$. In the renewal process, $J_m^n(l)$ is the $l$-th jumping time, or the time at which the $l$-th occurrence happens, and it has the following relation with the renewal process, $N_m^n(t)$:

$$N_m^n(t) = \sum_{l = 1}^\infty \mathbb{I}_{\{ J_m^n(l) \leq t \}} = \sup \{l : J_m^n(l) \leq t \}.$$
Another way to define $N_m^n(t)$ is as below:

\begin{algorithm}
\label{algo2}
\begin{algorithmic}[1]
\STATE \emph{Set $\tau = t_m^*$, $cntr = 0$, $N_m^n(t) = 0$}

\WHILE {\emph{$cntr < t$}}
	\IF{$\eta_m(\tau) = n$}
    	\STATE $cntr++$
    	\STATE $N_m^n(t) \ += S_m^n(\tau)$
    \ENDIF
    \STATE \emph{ $\tau++$}
\ENDWHILE
\end{algorithmic}
\end{algorithm}
By convention, $N_m^n(0) = 0$. 

In the following, we define another renewal process, $N_m(t)$:

$$N_m(t) = \sum_{u = t_m^*}^{t_m^* + t - 1} \Big ( \mathbb{I}_{ \{ S_m^1(u) = 1 \}} + \mathbb{I}_{ \{ S_m^2(u) = 1 \}} + \cdots + \mathbb{I}_{ \{ S_m^N(u) = 1 \}} \Big ).$$

Similarly, let $H_m(l)$ be the holding time (service time) of the $l$-th task after time slot $t_m^*$ receiving service from serve $m$, and call $\{H_m(l), l \geq 1\}$ the holding process. Then define $J_m(l) = \sum_{i = 1}^{l} H_m(l)$ for $l \geq 1$, and let $J_m(0) = 0$. In the renewal process, $J_m(l)$ is the $l$-th jumping time, or the time at which the $l$-th occurrence happens, and it has the following relation with the renewal process, $N_m(t)$:

$$N_m(t) = \sum_{l = 1}^\infty \mathbb{I}_{\{ J_m(l) \leq t \}} = \sup \{l : J_m(l) \leq t \}.$$

Note that the central scheduler makes scheduling decisions for server $m$ at time slots $\{ t_m^* + J_m(l), l \geq 1 \}$. We denote these scheduling decisions by $D_m(t_m^*) = \Big ( \eta_m(t_m^* + J_m(l)) : l \geq 1 \Big )$.

Consider the time interval $[t_m^*, t_m^* + T - 1]$ when $T$ goes to infinity. Define $\rho_m^n$ as the fraction of time that server $m$ is busy giving service to tasks that are $n$-local to this server, in the mentioned interval. Obviously, $\sum_{n = 1}^N \rho_m^n = 1$. Then equation \eqref{t_m^*} is followed by the following:

\begin{equation}
\begin{aligned}
& \lim_{T \rightarrow \infty} \frac{ \mathbb{E} \left [ \sum_{t = t_m^*}^{t_m^* + T - 1} \bigg ( \frac{S_m^1(t)}{\alpha_1} + \frac{S_m^2(t)}{\alpha_2} + \cdots + \frac{S_m^N(t)}{\alpha_N} \bigg ) \bigg | Z(t_0) \right ]}{T} \\
= & \lim_{T \rightarrow \infty} \Bigg \{ { \mathbb{E} \Bigg [ \mathbb{E} \Bigg [ \sum_{t = t_m^*}^{t_m^* + T - 1} \bigg ( \frac{S_m^1(t)}{\alpha_1} + \frac{S_m^2(t)}{\alpha_2} }\\
&{ \ \ \ \ \ \ \ \ \ \ \ \ \ \ \ \ \ \ \ \ \ \ \ \ \ \ \ + \cdots + \frac{S_m^N(t)}{\alpha_N} \bigg ) \bigg | D_m(t_m^*), Z(t_0) \Bigg ] \Bigg | Z(t_0) \Bigg ]} \Bigg \} \Bigg / T \\
= & \sum_{n = 1}^N \lim_{T \rightarrow \infty} \frac{ \mathbb{E} \Bigg [ \frac{1}{\alpha_n} \mathbb{E} \left [ \sum_{t = t_m^*}^{t_m^* + T - 1} \Big ( S_m^n(t)  \Big ) \bigg | D_m(t_m^*), Z(t_0) \right ] \Bigg | Z(t_0) \Bigg ]}{T} \\
= & \sum_{n = 1}^N \mathbb{E} \Bigg [ \frac{1}{\alpha_n} \lim_{T \rightarrow \infty} \frac{\mathbb{E} \left [ N_m^n \big ( \rho_m^n T \big ) \Big | D_m(t_m^*), Z(t_0) \right ]}{T} \Bigg | Z(t_0) \Bigg ].
\label{t_m^*2}
\end{aligned}
\end{equation}
Note that given $\{D_m(t_m^*), Z(t_0)\}$, the holding times $\{H_m^n(l), l \geq 1\}$ are independent and identically distributed with CDF $F_n$. If $\rho_m^n = 0$, then we do not have to worry about those tasks that are $n$-local to server $m$ since they receive service from this server for only a finite number of times in time interval $[t_m^*, t_m^* + T - 1]$ as $T \rightarrow \infty$, so
$$\lim_{T \rightarrow \infty} \frac{\mathbb{E} \left [ N_m^n \big ( \rho_m^n T \big ) \big | D_m(t_m^*), Z(t_0) \right ]}{T} = 0.$$
But if $\rho_m^n > 0$, we can use the strong law of large numbers for renewal process $N_m^n$ to conclude the following:

\begin{equation}
\lim_{T \rightarrow \infty} \frac{\mathbb{E} \left [ N_m^n \big ( \rho_m^n T \big ) \big | D_m(t_m^*), Z(t_0) \right ]}{T} = \rho_m^n \cdot \frac{1}{\mathbb{E}[H_m^n(1)]},
\label{t_m^*3}
\end{equation}
where the holding time (service time) $H_m^n(1)$ has CDF $F_n$ with expectation $\frac{1}{\alpha_n}$. Combining equations \eqref{t_m^*2} and \eqref{t_m^*3}, Lemma \ref{asymptoticT} is concluded as follows:

\begin{equation}
\begin{aligned}
& \lim_{T \rightarrow \infty} \frac{ \mathbb{E} \left [ \sum_{t = t_m^*}^{t_m^* + T - 1} \bigg ( \frac{S_m^1(t)}{\alpha_1} + \frac{S_m^2(t)}{\alpha_2} + \cdots + \frac{S_m^N(t)}{\alpha_N} \bigg ) \bigg | Z(t_0) \right ]}{T} \\
= & \sum_{n = 1}^N \mathbb{E} \left [ \frac{1}{\alpha_n} \cdot \rho_m^n \cdot \alpha_n \bigg | Z(t_0) \right ] = \sum_{n = 1}^N \rho_m^n = 1.
\label{t_m^*2}
\end{aligned}
\end{equation}

\subsection{Proof of Lemma \ref{lemma123456}}
\label{prooflemma123456}
Lemma \ref{lemma123456}:
Under the GB-PANDAS algorithm, for any arrival rate vector strictly inside the outer bound of the capacity region, $\mathbold{\lambda} \in \Lambda$, and any $\theta_1 \in (0, 1)$, there exists $T_1 > 0$ such that the following is true for any $T \geq T_1$:

\[
\begin{aligned}
& \mathbb{E} \Big [ ||\mathbold{\Psi}(t_0 + T)||_1 - ||\mathbold{\Psi}(t_0)||_1 \Big | Z(t_0) \Big ] \\
\leq & -\theta_1 ||\mathbold{\Psi}(t_0)||_1 + M T, \ \forall t_0 \geq 0,
\end{aligned}
\]
where $||.||_1$ is $L^1$-norm.

\textbf{Proof:}
For any server $m \in \mathcal{M}$, let $t_m^*$ be the first time slot after or at time slot $t_0$ at which the server is available ($t_m^*$ is also defined in \eqref{tm**}); that is,

\begin{equation}
t_m^* = \min\{ \tau: \tau \geq t_0, \Psi_m(\tau) = 0 \},
\end{equation}
where it is obvious that $\Psi_m(t_m^*) = 0.$ \\
Note that for any $t$, we have $\Psi_m(t + 1) \leq \Psi_m(t) + 1$, that is true by the definition of $\Psi(t)$, which is the number of time slots that server $m$ has spent on the currently in-service task. From time slot $t$ to $t+1$, if a new task comes in service, then $\Psi_m(t+1) = 0$ which results in $\Psi_m(t + 1) \leq \Psi_m(t) + 1$; otherwise, if server $m$ continues giving service to the same task, then $\Psi_m(t + 1) = \Psi_m(t) + 1$. Thus, if $t_m^* \leq t_0 + T$, it is easy to find out that $\Psi_m(t_0 + T) \leq t_0 + T - t_m^* \leq T$. In the following we use $t_m^*$ to find a bound on $\mathbb{E} [ \Psi_m(t_0 + T) - \Psi_m(t_0) | Z(t_0)]$:

\begin{equation}
\begin{aligned}
& \mathbb{E} \Big [ ||\mathbold{\Psi}(t_0 + T)||_1 - ||\mathbold{\Psi}(t_0)||_1 \Big | Z(t_0) \Big ] \\
= & \sum_{m = 1}^M \mathbb{E} \left [ \Big (\Psi_m(t_0 + T) - \Psi_m(t_0) \Big ) \bigg | Z(t_0) \right ] \\
= & \sum_{m = 1}^M \bigg \{ \mathbb{E} \left [ \Big (\Psi_m(t_0 + T) - \Psi_m(t_0) \Big ) \bigg | Z(t_0), t_m^* \leq t_0 + T \right ] \\
& \ \ \ \ \ \ \ \ \ \ \ \ \ \ \ \ \ \ \ \ \ \ \ \ \ \ \ \ \ \ \ \ \ \ \ \ \ \ \ \ \ \ \ \ \ \ \ \ \ \ \ \ \ \ \ \ \ \ \ \ \times P \left ( t_m^* \leq t_0 + T \big | Z(t_0) \right ) \\
& \ \ \ \ + \mathbb{E} \left [ \Big (\Psi_m(t_0 + T) - \Psi_m(t_0) \Big ) \bigg | Z(t_0), t_m^* > t_0 + T \right ] \\
& \ \ \ \ \ \ \ \ \ \ \ \ \ \ \ \ \ \ \ \ \ \ \ \ \ \ \ \ \ \ \ \ \ \ \ \ \ \ \ \ \ \ \ \ \ \ \ \ \ \ \ \ \ \ \ \ \ \ \ \ \times P \left (t_m^* > t_0 + T \big | Z(t_0) \right ) \bigg \} \\
\overset{(a)}{\leq} & \sum_{m = 1}^M \bigg \{ \Big (T - \Psi_m(t_0) \Big ) \times P \left (t_m^* > t_0 + T \big | Z(t_0) \right ) \\
& \ \ \ \ \ \ \ \ \ \ \ \ \ \ \ \ \ \ \ \ \ \ \ \ \ \ \ \ \ \ \ \ \ \ \ \ \ \ \ \ \ \ \ \ \ \ \ \ \ \ \ \ \ + T \times P \left (t_m^* > t_0 + T \big | Z(t_0) \right ) \bigg \} \\
= & - \sum_{m = 1}^M \bigg ( \Psi_m(t_0) \cdot P \left (t_m^* > t_0 + T \big | Z(t_0) \right ) \bigg ) + MT,
\label{sdfg}
\end{aligned}
\end{equation}
where $(a)$ is true as given that $t_m^* \leq t_0 + T$ we found that $\Psi_m(t_0 + T) \leq T$, so $\Psi_m(t_0 + T) - \Psi_m(t_0) \leq T - \Psi_m(t_0)$, and given that $t_m^* > t_0 + T$, it is concluded that server $m$ is giving service to the same task over the whole interval $[t_0, t_0 + T]$, which results in $\Psi_m(t_0 + T) - \Psi_m(t_0) = T$. \\
Since service time of an $n$-local task has CDF $F_n$ with finite mean, we have the following:

$$\lim_{T \rightarrow \infty} P \left (t_m^* \leq t_0 + T \Big | Z(t_0) \right ) = 1, \ \forall m \in \mathcal{M}$$
so for any $\theta_1 \in (0, 1)$ there exists $T_1$ such that for any $T \geq T_1$, we have $P \left (t_m^* \leq t_0 + T \Big | Z(t_0) \right ) \geq \theta_1,$ for any $m \in \mathcal{M}$, so equation \eqref{sdfg} follows as below which completes the proof:

\begin{equation}
\begin{aligned}
& \mathbb{E} \Big [ ||\mathbold{\Psi}(t_0 + T)||_1 - ||\mathbold{\Psi}(t_0)||_1 \Big | Z(t_0) \Big ] \\
\leq & - \theta_1 \sum_{m = 1}^M \Psi_m(t_0) + MT \\
= & -\theta_1 ||\mathbold{\Psi}(t_0)||_1 + M T.
\end{aligned}
\end{equation}

